\documentclass[11pt, a4paper]{article}
\pdfoutput=1

\usepackage[utf8]{inputenc}
\usepackage{jcappub}

\usepackage{amssymb,amsmath,latexsym,mathrsfs}
\usepackage{bm}
\usepackage{url}
\usepackage{epsfig,graphicx,verbatim,xspace,multirow,mathtools}
\usepackage{array}

\usepackage{mathtools}

\newcommand{\e}{{\rm ev}}

\newcommand{\keV}{\text{keV}}
\newcommand{\TN}{{T_{\rm ncdm}}}
\newcommand{\q}{\left(}

\newcommand{\w}{\right)}
\newcommand{\W}{{\rm WDM}}
\newcommand{\mpl}{M_p}

\title{Non-Cold Dark Matter from Primordial Black Hole Evaporation}

\author[a]{Iason Baldes,}
\emailAdd{iason.baldes@ulb.ac.be}
\author[a]{Quentin Decant,}
\emailAdd{quentin.decant@ulb.ac.be}
\author[a]{Deanna C. Hooper,}
\emailAdd{deanna.hooper@ulb.be}
\author[ab]{\\ and Laura Lopez-Honorez}
\emailAdd{llopezho@ulb.ac.be}

\affiliation[a]{Service de Physique Th\'eorique, CP225, Universit\'e Libre de Bruxelles, Bld du Triomphe, B-1050 Brussels, Belgium}

\affiliation[b]{Theoretische Natuurkunde, Vrije Universiteit Brussel and The International Solvay Institutes,
Pleinlaan 2, B-1050 Brussels, Belgium}

\abstract{Dark matter coupled solely gravitationally can be produced
  through the decay of primordial black holes in the early
  universe. If the dark matter is lighter than the initial black hole
  temperature, it could be warm enough to be subject to structure
  formation constraints. In this paper we perform a more precise
  determination of these constraints. We first evaluate the dark
  matter phase-space distribution, without relying on the
  instantaneous decay approximation. We then interface this
  phase-space distribution with the Boltzmann code {\sc class} to
  extract the corresponding matter power spectrum, which we find to
  match closely those of warm dark matter models, albeit with a
  different dark matter mass. This mapping allows us to extract
  constraints from Lyman-$\alpha$ data without the need to perform
  hydrodynamical simulations. We robustly rule out the possibility,
  consistent with previous analytic estimates, of primordial black
  holes having come to dominate the energy density of the universe and
  simultaneously given rise to all the DM through their
  decay. Consequences and implications for dark radiation and
  leptogenesis are also briefly discussed.}

\begin{document}

\hfill{\small ULB-TH/20-05}

\maketitle


\section{Introduction}
Dark matter (DM) is a cornerstone of both modern cosmology and
particle physics. Despite this, one of the most well-motivated
candidates, the Weakly Interacting Massive Particle (WIMP), in which
the DM is produced via freeze-out in the early universe, has thus far
avoided detection~\cite{Arcadi:2017kky}. Combined with possible
cosmological failings of the standard cold dark matter (CDM) paradigm
on small scales~\cite{Pawlowski:2015qta,Sawala:2015cdf,Tulin:2017ara},
this has reinvigorated interest in non-cold dark matter models
(NCDM)~\cite{Bode:2000gq}, and alternative production mechanisms
involving more feebly coupled DM candidates, see e.g.~\cite{Hall:2009bx,Chu:2011be,Bernal:2017kxu,Heeba:2018wtf}.

Another possibility is for DM to not be a particle, but rather a population of primordial black holes (PBHs). The early universe is thought to have undergone a period of inflation, diluting away any existing curvature, and along with it any matter or radiation. The inflaton field eventually decayed, reheating the universe, and imprinting its fluctuations on the radiation density. A large enough overdensity will collapse and form PBHs once the fluctuation re-enters the Hubble horizon~\cite{Carr:1993aq, Ivanov:1994pa,Kalaja:2019uju}. The PBHs, if massive enough to be stable on cosmological timescales, are a suitable DM candidate. This possibility, however, has been put under pressure from various considerations, including accretion and subsequent energy injection into the CMB~\cite{Ali-Haimoud:2016mbv,Poulin:2017bwe}, lensing observations~\cite{Alcock:1995dm}, and constraints from star formation~\cite{Capela:2012jz}, meaning much of the possible mass range is disfavoured~\cite{Bellomo:2017zsr}. Nevertheless, this remains an active area of research with many constraints having recently been either tightened or re-evaluated entirely~\cite{Carr:2009jm,Carr:2020gox, Carr:2020xqk}.

Alternatively, the PBHs may be light enough to decay via Hawking radiation~\cite{Hawking:1974rv,Hawking:1974sw} at an early enough epoch to avoid these constraints. Strong limits exist on PBHs which decay between big bang nucleosynthesis (BBN) and recombination from the observed yields of light elements~\cite{Kohri:1999ex,Carr:2009jm} and observations of CMB anisotropies~\cite{Poulin:2016anj,Stocker:2018avm,Poulter:2019ooo}, which may be further strengthened through the use of CMB spectral distortions~\cite{Lucca:2019rxf,Acharya:2019xla}. Early work on PBHs which decay before BBN focused on the possibility of the DM being composed of stable Planck scale remnants of the PBHs~\cite{MacGibbon:1987my, 
Barrow:1992hq, Carr:1994ar}, and has been extended in a number of ways~\cite{Dolgov:2000ht,Baumann:2007yr,Hooper:2019gtx, Dvali:2020wft}. Here we will instead assume the PBHs decay away completely, and consider particle DM candidates which -- even if they only have gravitational interactions -- can be produced through PBH decay~\cite{Matsas:1998zm,Bell:1998jk,Khlopov:2004tn,Fujita:2014hha,Allahverdi:2017sks,Lennon:2017tqq,Morrison:2018xla,Masina:2020xhk, Hooper:2020evu}. This results in a tight relation between the initial PBH mass, the DM mass, the initial density fraction of the universe in PBHs, and $T_\e$, the temperature of the universe immediately after the PBHs evaporate.

In this paper we revisit the production of NCDM lighter than the initial black hole temperature at formation and explore the different bounds on such models; coming from inflation, BBN, the observed DM relic abundance, the effective number of relativistic degrees-of-freedom $N_\mathrm{eff}$, and structure formation. For the first time, we will explore the full impact of the non-instantaneous evaporation of the PBHs on the resulting NCDM. To do this, we use the phase-space distribution of such models as an input for the Boltzmann code {\sc class}~\cite{Blas:2011rf}, which then allows us to extract the matter power spectrum, and thus the transfer function of these models compared to the standard $\Lambda$CDM case. This enables us in turn to constrain the model using the structure formation bounds from Lyman-$\alpha$ data~\cite{Viel:2005qj,Viel:2013fqw,Palanque-Delabrouille:2019iyz,Garzilli:2019qki}.

This paper is organised as follows. In Sec.~\ref{sec:black-holes-evap} we briefly introduce the context in which we consider PBH formation and their decay products. We go on in Sec.~\ref{sec:aevap} to find the DM mass required for the PBH decay to produce the observed DM relic abundance. In Sec.~\ref{sec:NCDM-PSD} we derive the resulting non-cold phase-space density of the DM. In Sec.~\ref{sec:lyman-alpha-n_eff} we constrain the scenario from its effects on structure formation by calculating the power spectrum and comparing it with Lyman-$\alpha$ limits. Possible contributions to $\Delta N_{\rm eff}$ are also discussed. The main results of our paper are presented in Sec.~\ref{sec:viable-param-space}, which summarises the allowed parameter space of the model. We then make some comments regarding the consequences for leptogenesis in Sec.~\ref{sec:lepto-from-pbh}, before concluding in Sec.~\ref{sec:concl}. We add further details on some of our calculations in the appendices. In particular, in App.~\ref{sec:greybody-factors} we discuss how a more detailed treatment of the greybody factors might affect our results, in App.~\ref{sec:valid-inst-evap} we address the validity of the instantaneous reheating assumption, in App.~\ref{ap: changing the DOF} we detail how changes in the number of degrees of freedom (dark or not) affect our results and, finally, in App.~\ref{sec:entropy} we derive the expressions for $\Delta N_{\rm eff}$

\section{Black holes: origin and evaporation}
\label{sec:black-holes-evap}
A non-rotating BH with zero charge and temperature given by\footnote{For a black hole in a stationary spacetime the temperature is defined as $T_{\mathrm{BH}} = \kappa/2\pi$, with $\kappa$ the surface gravity \cite{bardeen1973}. If the black hole is charged or rotating, its temperature will differ from eq.~\eqref{eq:TBH}.}
\begin{equation}
  T_{\mathrm{BH}}=\frac{M_p^2}{8 \pi M_{\mathrm{BH}}}\,,
  \label{eq:TBH}
\end{equation}
with the Planck mass $M_p= 1.22\times 10^{19}$\,GeV and $M_{\mathrm{BH}}$ the BH mass, will emit particles of the $j$th species at an averaged rate per energy interval
of:
\begin{equation}
  \frac {dN_j}{dt dE}=\frac{g_j}{2 \pi}\frac{\Gamma_j(E,M_{\mathrm{BH}})}{\exp\q E/T_{\mathrm{BH}}\w \pm 1} \,,
  \label{eq:dNdtdE}
\end{equation}
when considering fermion ($+$) or boson ($-$) emission, and $g_j$
denotes the number of degrees of freedom (dof) of the species $j$, see
e.g.~\cite{PhysRevD.41.3052, PhysRevD.13.198} for the more general
case. The coefficients $\Gamma_j(E, M_{\mathrm{BH}})$ are absorption
probabilities, referred to as {\it greybody factors}~\cite{Page:1976df}, which
depend on the energy of the particles, the mass of the BH, and the
properties of the particle $j$. They can be obtained by computing the
transmission coefficients of a wave of energy $E$ between the BH
horizon and spatial infinity, but they are non-trivial to evaluate
(see, however,~\cite{Arbey:2019mbc} for a new tool). They tend to
$\Gamma_j(E,M_{\mathrm{BH}})=27 E^2 {M_{\mathrm{BH}}^2}/{M_p^4}$ in the high energy (${E\gg T_{\mathrm{BH}}}$) geometrical-optics limit, while they fall off more
quickly as $E\to 0$, with higher spins producing stronger
cutoffs~\cite{Page:1976df}. Here in order to provide a self consistent
analysis and simple analytic estimates, we make use of the
geometrical-optics limit. We note, however, that this ignores small
spin-dependent low-$E$ suppressions of the spectrum, and as such will
lead to a slight underestimation of the total portion of relativistic
(high-$E$) particles as already mentioned in~\cite{Lennon:2017tqq}. We
provide more details on the expected effects of a detailed processing
of greybody factors in App.~\ref{sec:greybody-factors}.

Within this context, combining eqs.~(\ref{eq:dNdtdE}) and
(\ref{eq:TBH}), the emitted power implies that the BH mass decreases
with time at a rate
\begin{equation}
  \frac {dM_{\mathrm{BH}}}{dt }=-\sum_j\int_{0}^{\infty} E \frac {dN_j}{dt dE} dE= - e_T\frac{M_p^4}{M^2_{\rm BH}}\,,
   \label{eq:dMdt}
\end{equation}
where $e_T$ is given by
\begin{equation}
  e_T=\frac{27}{4} \frac{ g_{* \mathrm{BH}}}{30720 \pi} \, .
  \label{eq:eT}
\end{equation}
Here we have introduced $g_{* \mathrm{BH}}$, the total number of
relativistic degrees of freedom (i.e. with $m_j<T_{\mathrm{BH}}$)
emitted by the BH. For definitiveness, in the following we will assume
that the BH will emit a two-component fermionic DM particle with mass
$m_{\mathrm{DM}}< T_{\mathrm{BH}}$ together with Standard Model (SM)
particles.  In particular, for BHs with a temperature above the approximate
electroweak phase transition temperature, $T_{\rm EW}$, the SM
relativistic dof plus a two-component fermionic DM particle give rise
to
\begin{equation}
  g_{* \mathrm{BH}}=108.5\quad {\rm}\quad e_T=7.6\times 10^{-3}\quad [T_{\rm BH} >T_{\rm EW}] \,,
\label{eq:eTnum}
\end{equation}
see App.~\ref{sec:greybody-factors} for more details. The effects of
changing the number of dof, are explored in detail in App.~\ref{ap:
  changing the DOF}.

From eq.~(\ref{eq:dMdt}), the BH mass evolves with time as follows:
\begin{equation}
  M_{\mathrm{BH}}(t)= M_F\left(1- \frac{(t-t_F)}{\tau}\right)^{1/3}\,,
  \label{eq:MBH}
\end{equation}
with $M_F$ the BH mass at formation, and $t_F$ the time of formation. The BH lifetime, $\tau$, reads
\begin{equation}
  \tau=\frac{1}{3e_T} \frac{M_F^3}{M_p^4}\,.
  \label{eq:tau}
\end{equation}
Furthermore, by integrating eq.~(\ref{eq:dNdtdE}) over energy and time, we can
also compute the total number of particle species $j$ emitted over the
PBH lifetime. In particular, for a fermionic species $j$, we have
\begin{eqnarray}
  N_j&=& g_j\frac{ 81\zeta(3)}{4096 \pi^4 e_T}\frac{M_F^2}{M_p^2}= 3.2 \times 10^{-2} g_j \frac{M_F^2}{M_p^2} \quad [j\equiv{\rm fermion}]\, ,
\label{eq:Nj}
\end{eqnarray}
where we have used $e_T$ from eq.~(\ref{eq:eTnum}) in the second equality.

The question that actually arises is what should we expect as initial
mass $M_F$ and formation time $t_F$. Here we assume that the inflaton
decays into radiation with an overdensity on a suitably small
scale. If this overdensity is larger than the equation of state 
parameter, $w=1/3$, it will collapse into a PBH of mass
\begin{equation}
  M_F=\gamma \rho_F\frac{4\pi}{3} H_F^{-3} \,,
  \label{eq:HF}
\end{equation}
where $H_{ F}=1/(2t_{\rm F})$ is the Hubble scale at PBH formation in a
radiation era, and ${\gamma \sim w^{3/2} \approx 0.2}$ captures the
efficiency of collapsing the overdense region into the PBH~\cite{Carr:1975qj}. We
can now use the Friedmann equation $H_F^2= 8\pi \rho_F/(3 M_p^2)$ to
relate $H_F$ to the total energy density at formation time,
$\rho_F$,
\begin{equation}
	\rho_F = 3\q 4\pi\gamma\w^2 M_F T_F^3\,, 
\label{eq:rhof}
\end{equation}
where $T_F = M_p^{2}/(8\pi M_{F})$ is the BH temperature at formation
time.
Hence, the time of formation is given by 
\begin{equation}
	t_F = \frac{M_F}{\gamma \mpl^2} \,.
\label{eq:formationtime}
\end{equation}
The constraint on the tensor-to-scalar ratio from the CMB limits the scale of inflation to $H_{\rm Inf} \lesssim 10^{14}$\,GeV~\cite{Akrami:2018odb}, and in turn, as $H_{\rm F} < H_{\rm Inf}$, the initial PBH mass should satisfy the lower bound:
\begin{equation}
  M_F \gtrsim  10^4 M_p\,.\quad{\rm [Inflation]}
  \label{infl}
\end{equation}

\noindent Following the literature, we denote the initial PBH abundance as
\begin{equation}
  \beta \equiv  \Omega_{\rm PBH}(t_{ F}) \,,
\end{equation}
which allows us to express the initial PBH number density as
\begin{eqnarray}
  n_{\mathrm{BH}}(t_F)&=&\frac{\beta}{ M_F}\rho_F= 3 \beta (4 \pi \gamma)^2 T_F^3 \,.
\label{eq:nBHtF}
\end{eqnarray}
As fluctuations of the density contrast which exceed the threshold value $w = 1/3$ and collapse can be rare, not all Hubble patches at $t_{F}$ will produce a PBH, and it is possible to have $\beta \ll 1$~\cite{Morrison:2018xla}. 
After being produced in the radiation dominated era, the energy density of PBHs initially grows as
	\begin{equation}
	\Omega_{\rm BH} (t)= \frac{ \rho_{\rm BH} (t)}{ \rho_{\rm tot}(t) } \propto a (t)\,,
	\end{equation}
where $\rho_{\rm tot}$ is the total density, originally dominated by the
radiation density $\rho_R$, and $a$ is the scale factor. Provided they
do not decay beforehand, we reach $\rho_{\rm BH}/\rho_R \approx 1$
at the PBH-radiation equality time, $t_{\rm eq}$. Assuming that $a(t)\propto
t^{1/2}$ up to equality, we have
\begin{equation}
  t_{\rm eq}\simeq t_F/\beta^2\,.
  \label{eq:teq}
\end{equation}
In order for BHs to dominate the universe before
evaporation -- thus giving rise to an early matter dominated era -- $t_{\rm eq}$
should be smaller than the evaporation time, $t_\e=t_F+\tau\simeq
\tau$ (this is valid for the cases considered here, see
below). Equivalently, we need $\beta>\beta_c$ with
\begin{equation}
         \beta_c=\sqrt{\frac{3e_T}{\gamma}}\frac{M_p}{M_F}\,.
\label{eq:betac}
\end{equation}
Note that the temperature of the plasma surrounding the PBH at formation,
	\begin{equation}
	T(t_{F}) = \left(\sqrt{ \frac{45}{16 \pi^{3} g_*(t_{F}) } } \frac{\gamma M_{p}^{3} }{ M_F } \right)^{1/2},
	\label{eq:plasmaTatform}
	\end{equation}
exceeds the BH temperature $T_{F}$. Hence, the PBHs do not start decaying immediately, but only at some later time, either when the plasma has cooled sufficiently, or when the PBHs grow to dominate the energy density, depending on which occurs first. In either case $t_\e$ is still dominated by the PBH lifetime $\tau$. The estimate of the efficiency of the Bondi accretion~\cite{Bondi:1952ni} onto the PBHs carries a large uncertainty, see e.g.~\cite{Carr:1974nx,Custodio:2002gj,Guedens:2002sd}. Moreover, even if the accretion efficiency is close to unity, the corrections to the ``initial" BH mass and $\tau$ are only of $\mathcal{O}(1)$~\cite{Hooper:2019gtx,Masina:2020xhk}. We therefore do not include such effects in our study but keep in mind the possibility of some corrections if the accretion is indeed efficient. 

Here we note that we have thus far made the assumption that the standard description of Hawking radiation is valid for any of the 
BH masses considered in this work, meaning that the BHs fully evaporate, leaving no remnant behind. However, some effects could  halt or slowdown this process and give rise to the possibility of smaller compact objects, which could account for part or all of the DM~\cite{MacGibbon:1987my, 
Barrow:1992hq, Carr:1994ar, Dolgov:2000ht,Baumann:2007yr,Hooper:2019gtx, Dvali:2020wft}. If this is indeed the case, our analysis below would, of course, have to be modified.

\section{Relic dark matter from evaporation}
\label{sec:aevap}

Sourcing the DM relic abundance from BH decay has been considered in a number of previous studies~\cite{Matsas:1998zm,Bell:1998jk,Khlopov:2004tn,Fujita:2014hha,Allahverdi:2017sks,Lennon:2017tqq,Baumann:2007yr,Morrison:2018xla,Hooper:2019gtx,Masina:2020xhk, Hooper:2020evu}. Given an initial BH mass, $M_{F}$, there generally exist two solutions to match onto the relic abundance, depending on whether the initial BH temperature, $T_{F}$, is above or below $m_{\rm DM}$, see e.g.~\cite{Fujita:2014hha,Lennon:2017tqq}. In the case of $m_{\rm DM} > T_{F}$, solutions exist with $m_{\rm DM} < M_{p}$ for monochromatic mass functions with $M_{F} \gtrsim 10^9 M_{\rm P}$. The DM ends up cold enough to not be subject to structure formation constraints~\cite{Fujita:2014hha} and will be of no further interest to us here.  

We now proceed to estimate the relic abundance of light DM, $m_{\rm DM} < T_{F}$, from PBH evaporation. Barring additional non-standard expansion, from evaporation time onward the DM rest energy density, $n_{\mathrm{DM}}m_{\mathrm{DM}}$, scales as $a^{-3}$, meaning that\footnote{Notice that light
  DM produced by PBH evaporation is relativistic at the time
  of evaporation and a large part of its total energy density is kinetic
  energy. However, the latter quantity redshifts fast enough in order to  be neglected today, see Sec.~\ref{sec:delta-n_eff} for a detailed analysis.}
\begin{equation}
\Omega_{\mathrm{DM}}(t_0)= \frac{m_{\mathrm{DM}} n_{ \rm DM}(t_\e)}{\rho_c} \times \left( \frac{ a_\e }{ a_0 }\right)^3 \,,
\label{eq:omdm0}
\end{equation}
where $n_{ \rm DM}(t_\e)$ is the DM number density at evaporation, $\rho_c$ is the critical energy density today, and $a_0 \equiv 1$ is the scale factor today. We can write $n_{ \rm DM}(t_\e) = N_{\mathrm{DM}} n_{\mathrm{BH}} (t_\e)$, where $n_{\mathrm{BH}} (t_\e)$ is the BH number density at evaporation time and $N_{\mathrm{DM}}$ is the total number of DM particles emitted by one PBH. This latter quantity is given in eq.~(\ref{eq:Nj}) for fermionic DM.

Let us first evaluate the scale factor at evaporation. We use entropy conservation between evaporation and today\footnote{The entropy is not necessarily conserved between PBH formation and evaporation: if $\beta > \beta_c$ the new relativistic plasma arising from the PBH gives rise to a non-negligible increase of entropy, see e.g.~\cite{Chaudhuri:2020wjo}.}, i.e.~$ s_0 a_0^3=s_\e a_\e^3$, where $s_0$ denotes the present day entropy density. Immediately after evaporation the universe is radiation dominated, meaning that
\begin{equation}
  H^2 =\frac{8\pi }{3M_p^2} \rho_R(t_\e)= \frac{8 \pi^3 g_*(t_\e) T^4_\e}{90 \, M_p^2}   \,,
  \label{eq:H-Tev1}
\end{equation}
where $\rho_R(t_\e)$ and $T_\e$ are the radiation energy density and the temperature at evaporation. Now, depending on $\beta$, the universe was either matter dominated (MD) or radiation dominated (RD) before evaporation, giving rise to two different evaporation scale factors. The Hubble scale at evaporation is $H_\e =1/(2t_\e)$ for RD or $H_\e = 2/(3t_\e)$ for MD. In either case, $T_\e\propto 1/\sqrt{t_\e}$. Then using $ s_0 a_0^3=s_\e a_\e^3$, we obtain for the RD ($\beta < \beta_{c}$) scale factor 
\begin{align}
  a_\e^{\rm RD} = & \;  \left(\frac{45 s_0}{2\pi^2g_{*s}(t_\e)}\right)^{1/3} \left(\frac{32\pi^3g_*(t_\e)t_\e^{2}}{90 M_p^2}\right)^{1/4} \nonumber \\ = & \; 2.5\times 10^{-31} \left(\frac{M_{F}}{M_p}\right)^{3/2} \label{eq:aRD}.
\end{align}
where $g_{*s}$ counts the effective entropic degrees of freedom, see App.~\ref{sec:entropy}. The second equality is valid for $T_\e>T_{\rm EW}$. For the MD case ($\beta > \beta_{c}$) we instead find a slightly different result:
\begin{equation}
  a_\e^{\rm MD} =  \left(\frac{9}{16}\right)^{1/4} \, a_\e^{\rm RD}\, = 2.2\times 10^{-31} \left(\frac{M_{F}}{M_p}\right)^{3/2}\,.\label{eq:aMD}
\end{equation}

Additionally, we can evaluate the PBH number density at evaporation
time $n_{\mathrm{BH}}(t_\e)$ as a function of the PBH number density
at formation, eq.~(\ref{eq:nBHtF}), by determining the ratio of scale
factors between evaporation and formation. Here, we assume that the
number of dof does not change between PBH formation and
evaporation. We relax this assumption in App.~\ref{ap: changing the
  DOF}.  The simplest case is when the universe is dominated by
radiation the whole time. Then the scale factor simply scales as
$a(t)\propto t^{1/2}$ and
\begin{equation}
  \frac{a(t_F)}{a(t_\e)}= \left(\frac{t_F}{t_\e}\right)^{1/2} =\left( \frac{ 3e_T}{\gamma}\right)^{1/2}\frac{M_p}{M_F}\quad {\rm if }\quad \beta<\beta_c\,.
\label{eq:RaRD}
\end{equation}
Alternatively, in the case where BHs can dominate the universe, the $a(t)\propto t^{1/2}$ dependence cannot be expected at all times between $t_F$ and
$t_\e$. A good estimate of the ratio can instead be obtained by assuming that
$\rho_{\mathrm{BH}}(t_\e)\approx\rho_R(t_\e)$ and that $\rho_{\mathrm{BH}}\propto a^{-3}$ between $t_F$ and $t_\e$, effectively neglecting the loss in mass of the BH by assuming that it happens rather instantaneously around $t\simeq t_\e$. Using that $\rho_R(t_\e)=3 M_p^2/(8\pi)\times (2/(3t_\e))^2$, we can easily extract the following expression:
\begin{equation}
  \frac{a(t_F)}{a(t_\e)} = \left(\frac{ 16 e_T^2}{\gamma^2\beta}\frac{M_p^4}{M_F^4}\right)^{1/3} \quad {\rm if }\quad \beta>\beta_c\,.
  \label{eq:RaMD}
\end{equation}
Combining eq.~(\ref{eq:omdm0}) with eqs.~(\ref{eq:RaRD})-(\ref{eq:RaMD}), we find
\begin{eqnarray}
  \Omega_{\mathrm{DM}}(t_0) &=&\frac{m_{\mathrm{DM}} N_{\mathrm{DM}} T_F^3}{\rho_c}\times \zeta_{\rm RD/MD}\times \q a_\e^{\rm RD/MD}\w^3 \,,
  \label{eq:OmDM1}
  \end{eqnarray}
where $\zeta_{\rm RD/MD}$ is a RD or MD dependent reheating (RH) prefactor which reads
\begin{equation}
	\zeta_{\rm RD/MD} \simeq   \begin{dcases*}
    3 \, \beta \, (4\pi)^2 \, \gamma^{1/2} \, (3e_T)^{3/2} \, \frac{M_p^3}{M_F^3} & if  $\beta<\beta_c$\,,\\
   3 \, (4\pi)^2 \, (4e_T)^{2} \, \frac{M_p^4}{M_F^4}&  if   $\beta>\beta_c$\,.
\end{dcases*}
  \label{eq:prefMDRD}
\end{equation}
Equivalently, using eqs.~(\ref{eq:aRD})-(\ref{eq:aMD}) with $T_\e>T_{\rm EW}$, we have:
\begin{equation}
  \frac{ \Omega_{\mathrm{DM}}(t_0)h^2 }{ 0.12 }  = \left(\frac{m_{\mathrm{DM}} }{ 1\, \rm  MeV}\right) \times
  \begin{dcases*}
   \left(\frac{ M_F}{1.1 \times 10^7 M_p }\right)^{1/2} \left( \frac{\beta}{3.6\times 10^{-8}}\right)  & if  $\beta<\beta_c$\,,\\
   \left(\frac{ M_F}{ 1.1 \times 10^7 M_p }\right)^{-1/2} &  if   $\beta>\beta_c$\,.
  \end{dcases*}
\label{eq:OmDM2}
\end{equation}
From the above result, it is clear that the DM abundance scales very differently with the initial PBH mass $M_F$ depending if their initial density $\beta$ is larger or smaller than the critical density. For $T_\e< T_{\rm EW}$, the number of dof $g_{*}(t_\e)$ decreases and slightly affects both $a_\e$ and $\Omega_{\rm DM}$. For a detailed treatment, see App.~\ref{ap: changing the DOF}.

During the early matter dominated epoch, overdensities, defined as $\delta \equiv \rho/\rho_{\rm tot} -1$, will grow linearly with the scale factor if they are subhorizon scale. We assume such overdensities, at length scales much larger than the initial overdensity which lead to PBH formation, exist at a CMB-inspired level of $\delta_i \sim 10^{-5}$. These overdensities will grow until the PBHs decay into radiation, which includes our DM, whereupon the overdensity is efficiently suppressed provided it is in the linear regime~\cite{Erickcek:2011us,Barenboim:2013gya,Miller:2019pss}. Indeed, the decay into radiation leads to oscillations in the plasma which can be a source of gravitational waves if the PBH mass function is sufficiently narrow~\cite{Inomata:2020lmk}. Therefore, we do not need to be concerned about additional structure formed due to the period of early matter domination as long as the overdensities do not enter the non-linear regime before evaporation. If they enter the non-linear regime, additional long-living structures such as larger PBHs could presumably be formed, although the dynamics are non-trivial. As our analyses do not take into account such non-linearities, the range of validity entails
\begin{equation}
\frac{ \delta_\e }{ \delta_i} = \frac{ a(t_\e) }{ a(t_{\rm eq}) } \approx \left( \frac{ \beta }{ \beta_c } \right)^{4/3} < 10^5\,, 
\end{equation}
where $\delta_\e$ ($\delta_i$) is the overdensity at evaporation (initial overdensity). In deriving the above we have used eq.~\eqref{eq:teq}, which implies $a(t_F)/a(t_{\rm eq})\approx \beta$, and eq.~\eqref{eq:RaMD}. Translating the above for convenience, our analysis will be valid for 
\begin{equation}
 \beta \lesssim 6 \times 10^{3} \beta_{c} \,.
\label{eq:limnon-lin}
\end{equation}

Independently of DM production and structure formation, we have to demand that the BHs evaporate before BBN, in order not to spoil agreement with observations~\cite{Kohri:1999ex,Carr:2009jm}. The precise limit depends on $\beta$ but becomes very stringent below ${T_\e \sim 4 \, {\rm MeV}}$. In order to be safe, we impose a simplified constraint, that the temperature at evaporation $T_\e> 10 \, {\rm MeV} > T_{\rm BBN}$. Making use of the Hubble rate and evaporation scale factors derived in this section, this translates into an upper bound on the initial BH mass,
\begin{equation}
M_F \lesssim 1.6\times 10^{13} M_p \quad \text{[BBN]}\,,
\label{eq:BBN}
\end{equation}
where we have used the number of relativistic dof at $T_\mathrm{ev}=10$\,MeV, $g_{*} = 10.75$.

\section{Non-cold dark matter phase-space  distribution}
\label{sec:NCDM-PSD}
Light DM particles ($m_{\mathrm{DM}}\ll T_{\mathrm{BH}}$) emitted ultra-relativistically from PBH evaporation might leave a NCDM~\cite{Lesgourgues:2011rh,Murgia:2017lwo} imprint on cosmological observables. In order to estimate this effect, we study in more detail the form of the momentum distribution. Here we will consider that the energy of the emitted particles is momentum dominated ($E\simeq p$). Using eq.~(\ref{eq:dNdtdE}) in this limit, and integrating over time
between $t_F\simeq 0$ and $\tau$, we can obtain the momentum distribution of particles of the species $j$ arising from one BH at the time of evaporation, $dN_j/dp|_{t=t_\e}$.

A priori, one could distinguish two different scenarios, one in which
the reheating would happen instantaneously after production,
i.e. $t_F\sim \tau$, and a non-instantaneous reheating, in which case
we should account for the redshifting of momenta of the emitted
particles between $t_F$ and $\tau$ with $\tau\gg t_F$. Given the lower
bound on the BH mass from inflation, in eq.~(\ref{infl}), we can
easily check that the range of allowed BH masses always forces us to
take into account the non-instantaneous reheating case, see
App.~\ref{sec:valid-inst-evap} for details.  Nonetheless, the case
of instantaneous reheating allows to extract exact analytic
results. Here, for the sake of completeness, we revisit the resulting
velocity distribution for both cases in
subsections~\ref{sec:inst-rehe} and \ref{sec:extended-reheating}. Our
results fully agree with previous findings of~\cite{Lennon:2017tqq},
see Fig.~\ref{fig:spectr}.

Additionally we go one step further by deriving the resulting DM phase-space
distribution, see Sec.~\ref{sec:ncdm-phase-space}, and interfacing it
with a Boltzmann code, so as to extract a more precise imprint on the
linear matter power spectrum, see
Sec.~\ref{sec:lyman-alpha-n_eff}. The public code {\sc
  class}~\cite{Blas:2011rf,Lesgourgues:2011rh} allows the treatment of
NCDM and involves the introduction of a NCDM
temperature\footnote{Notice that the parameter $\tt T_{\rm ncdm}$ of
  {\sc class} is effectively a ratio of temperatures. In terms of the
  variables defined here, ${\tt T_{\rm ncdm}}=\TN(t)/T(t_0)\times
  a(t)$, where $T(t_0)$ is the radiation temperature today. Let us
  emphasize that $\TN(t)$ is time dependent while  ${\tt T_{\rm ncdm}}$ is not. }
\begin{equation}
\TN(t)=T_F\frac{a(t_\e)}{a(t)} \,,  
\label{eq:Tncdm}
\end{equation}
such that $T_F=\TN(t_\e)$. We also introduce the time-independent rescaled momentum variable
\begin{equation}
  x(t)= \frac{p(t)}{\TN(t)}\,,
  \label{eq:x}
\end{equation}
which is nothing but the rescaled comoving momentum $\tt q$ of {\sc class} for NCDM particles. With these variables, we write the rescaled momentum distribution $dN_j/dx= g_j\, \xi\times  \tilde f(x)$ with
\begin{equation}
\tilde f(x) = \frac{T_F^3} {M_p^2g_j}\left.\frac{dN_j}{dp}\right|_{t=t_\e} \,, \label{eq:tildf0}
\end{equation}
where $\tilde f(x)$ is a universal momentum distribution, which is independent of the PBH mass, as will become clear in the next section, and the dimensionless coefficient is given by
\begin{equation}
  \xi= \frac{M_p^2}{T_F^2}\,.
  \label{eq:xi}
\end{equation}
Furthermore, in order to provide an estimate of the typical mean velocity of the particle species $j$ obtained from BH evaporation at evaporation
time $t=t_\e$, we will evaluate
\begin{equation}
  \langle p_j \rangle_{t=t_\e}=\frac{\int dx \, x\times T_F \,\tilde f(x)}{\int dx  \,\tilde f(x)}\,.
  \label{eq:meanp}
\end{equation}

\subsection{Instantaneous reheating}
\label{sec:inst-rehe}

When considering instantaneous reheating, we expect that the particle momenta do not have time to be redshifted between the time of production and the end of reheating, so that the momentum distribution of particles of species $j$ arising from a PBH reads
\begin{equation}
  \left.\frac{dN_j}{dp}\right|_{t=t_\e}=\int_{t_F}^{t_\e}\frac{dN_j}{dp dt' }(p,t') dt'\,.
  \label{eq:dNdpIns}
\end{equation}
In practice, we know that $t_F\ll t_\e$ for the BH mass range of interest. In this context, integrating the instantaneous distribution of eq.~(\ref{eq:dNdpIns}) from $t_F\simeq 0$ to $t_\e\simeq \tau$, we can rewrite the momentum distribution for e.g. a fermionic species in terms of the universal distribution function
\begin{eqnarray}
  \tilde f(x)&=&\frac{F_0}{ x^{3}}\int_{0}^{x}\frac{y^4}{\exp(y)+1 } dy \quad {\rm with}\quad  F_0=\frac{15 }{8\pi^5g_{* \mathrm{BH}}}\,.
\label{eq:dNdpAnal}
\end{eqnarray}
This function is manifestly independent of the BH mass. Also notice that the above integral can be
expressed in terms of special functions:
\begin{eqnarray}
  \int_{0}^{x}\frac{y^4}{\exp(y)+1 } dx&=&\frac{x^5}{5}-x^4 \log(1+e^{x})- 4 x^3 {\rm Li}_2 (-e^{x})+ 12 x^2 {\rm Li}_3 (-e^{x})\cr&&- 24 x {\rm Li}_4 (-e^{x})+24 {\rm Li}_5 (-e^{x})+ \frac{45}{2}\zeta(5) \,,
  \label{eq:dNdpAnal2}
\end{eqnarray}
where ${\rm Li}_n$ are polylogarithms.

\begin{figure*}[t]
  \begin{center}
  \includegraphics[width=.78\textwidth]{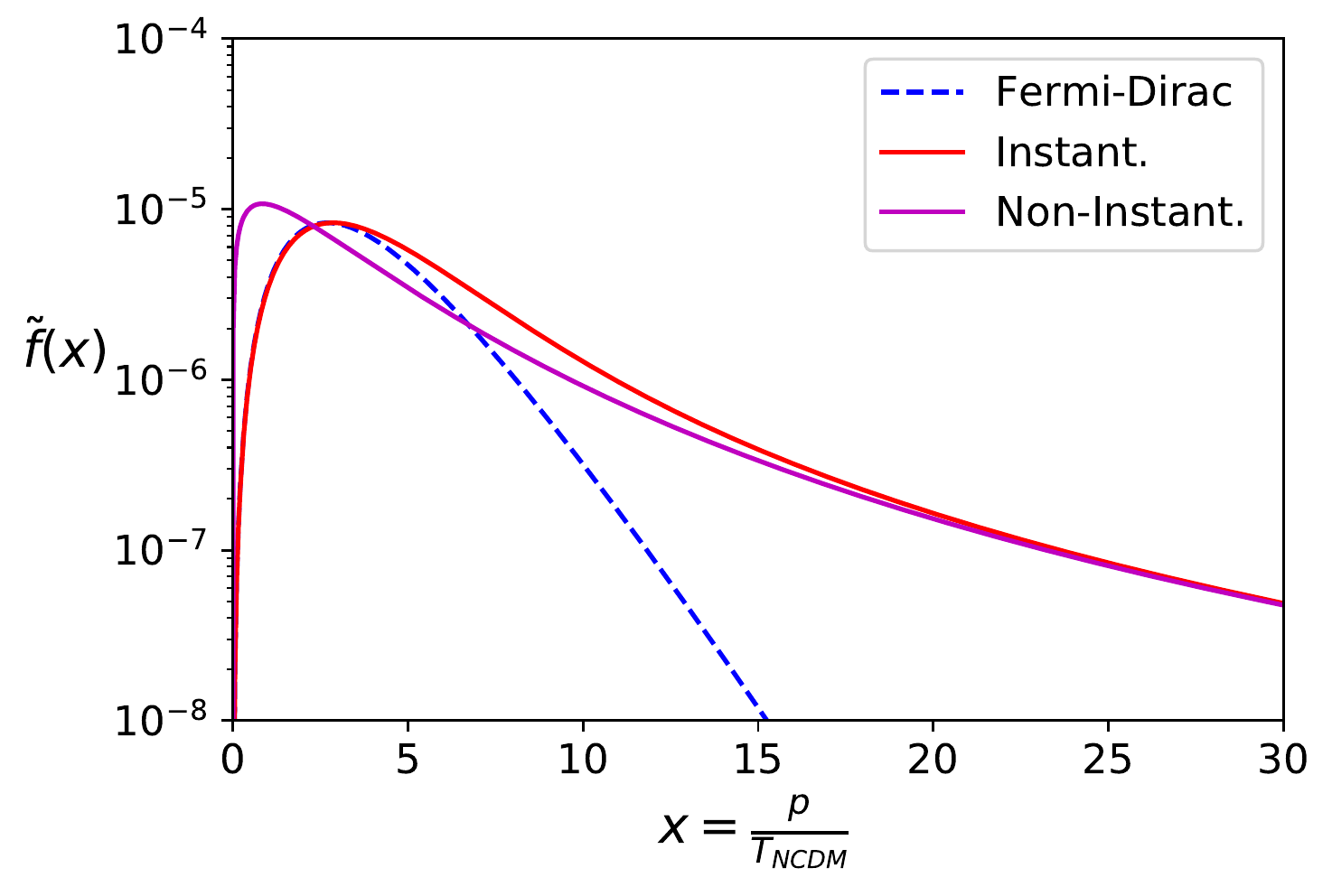}
  \vspace*{-5mm}
  \end{center}
\caption{The continuous red and purple curves depict the universal
  NCDM momentum distribution $\tilde f (x)$ of eq.~(\ref{eq:tildf0})
  arising from PBH evaporation as a function of the rescaled momentum
  $x$, assuming instantaneous and non-instantaneous reheating respectively. The
  dashed blue curve corresponds to a rescaled Fermi-Dirac distribution
  with a thermal body temperature $\tilde T=1.3 \times T_F$. See
  Sec.~\ref{sec:inst-rehe} and Sec.~\ref{sec:extended-reheating} for
  details.}
\label{fig:spectr}
\end{figure*}
The universal distribution of eq.~(\ref{eq:dNdpAnal}) for instantaneous reheating is shown as a function of the momentum $x$ in Fig.~\ref{fig:spectr}
with a red continuous line.  One can compare this distribution to a Fermi Dirac phase-space distribution $f_{\rm FD}$. For reference, we also show the renormalised distribution $N_{\rm norm} (p/\tilde T)^2 f_{\rm FD}(p,\tilde T)$ with a blue dashed curve in Fig.~\ref{fig:spectr}. The thermal body temperature $\tilde T=1.3 \times T_F$ and the normalisation factor $N_{\rm norm}$ were chosen so as to get the peak of the distributions at the same position. The thermal distribution and the distribution arising from PBH evaporation match very closely at low momenta, in particular around the position of
its maximum. However, in the case of PBH evaporation we get a higher velocity
tail than the distribution arising from a thermal body of temperature
$1.3 \times T_F$.  We can obtain an analytic form of the mean momentum
at evaporation from instantaneous PBH evaporation using eq.~(\ref{eq:meanp}):
\begin{equation}
  \langle p_j \rangle|_{t=\tau}=\frac{14\pi^3}{1440 \zeta(3)} \frac{M_p^2}{M_F}\approx 6.3\times  T_F\,,
  \label{eq:meanpEvap}
\end{equation}
in agreement with~\cite{Morrison:2018xla}.

This can be compared to the mean momentum resulting from a FD thermal
distribution. In the latter case, we have $\langle
p_j\rangle_{FD}=\frac{7\pi^4}{180 \zeta(3)} \tilde T \approx 3 \tilde T$, where
$\tilde T$ is the thermal body temperature, i.e. the mean momentum of
NCDM arising from a PBH is higher than the one of thermal NCDM.

\subsection{Non-instantaneous reheating}
\label{sec:extended-reheating}
In the case of  a non-instantaneous reheating, in which case momenta can redshift between the initial time and the time of evaporation, the momentum distribution can be computed using
\begin{eqnarray}
  \left.\frac{dN_j}{dp}\right|_{t=t_\e}
     &=&\int_{0}^{\tau}dt' \,\frac{a(\tau)}{a(t')}\times \frac{dN_j}{dp'  dt' }\left(p\frac{a(\tau)}{a(t')} ,t'\right) \,.
  \label{eq:dNdpslow}
\end{eqnarray}
A priori, one should solve the coupled set of Boltzmann equations giving rise to the Hubble rate dependence in $\rho_{\mathrm{BH}}$ and $\rho_R$ so as to extract the time dependence of $a(t)$. However, assuming that the reheating period is dominated by a fluid of equation of state $w$ throughout, the scale factors
will scale as $ a(t)=t^{2/(3(1+w))}$. One can check that considering $w=0$ or 1/3 induces only very small changes in the momentum distribution. Hence, we will use this approximation instead of solving the full system.

In Fig.~\ref{fig:spectr} we show with a purple continuous curve the resulting universal distribution function assuming a MD era before RH for a fermionic species. It presents a peak at lower velocities than in the instantaneous case, while the high velocity tail is unaffected, as already observed in~\cite{Lennon:2017tqq}. However, the area under the red and purple curves is the same, meaning that the number of particles emitted by one BH (obtained in eq.~(\ref{eq:Nj})) is independent of the fast/slow reheating, as one would expect. We can further extract the typical mean momentum associated to the non-instantaneous RH. For a fermionic species we get a mean momentum at
evaporation of
\begin{equation}
\langle p_j \rangle|_{t=\tau}\approx T_F \times 
\begin{dcases*}
5.3 & for  $\beta<\beta_c$\,, \\
5.1 & for  $\beta>\beta_c$\,,
\end{dcases*}
\label{eq:meanpEvapRDMD}
\end{equation}
i.e. (slightly) lower values than in the instantaneous case of eq.~(\ref{eq:meanpEvap}), even though the peak in the velocity distribution appears at a momentum lower by a factor of $\sim 4$ in the non-instantaneous reheating case.

\subsection{NCDM phase-space distribution }
\label{sec:ncdm-phase-space}

We define the DM phase-space distribution $f_{\mathrm{DM}}$ as
\begin{equation}
  g_{\mathrm{DM}} f_{\mathrm{DM}}(p,t)= \frac{dn_{\mathrm{DM}}}{d^3p}\,,
\end{equation}
where $g_{\mathrm{DM}}$ is the number of DM degrees of freedom; $n_{\mathrm{DM}}$ is the
DM number density, scaling as $a^{-3}$ from evaporation time; and $p$ is
the momentum, scaling as $1/a$. The $f_{\mathrm{DM}}$ arising from a
distribution of PBHs that is peaked on a given BH mass can be expressed
in terms of the momentum distributions derived above as
\begin{equation}
  f_{\mathrm{DM}}(p,t) d\Omega= \frac{1}{g_{\mathrm{DM}}}\frac{ n_{\mathrm{BH}}(t)}{ p(t)^3} x\left.\frac{dN_{\mathrm{DM}}}{dx}\right|_{t=t_\e} \,,
  \label{eq:fDM}
\end{equation}
where we have defined $n_{\mathrm{BH}}(t)=n_{\mathrm{BH}}(t_F)(a(t_F)/a(t))^3$. Making use of the BH number density at formation time of eq.~(\ref{eq:nBHtF}), of the ratio of scale factors between formation and evaporation time of eqs.~(\ref{eq:RaRD})-(\ref{eq:RaMD}), and the universal momentum distribution of eq.~(\ref{eq:tildf0}), we can rewrite eq.~(\ref{eq:fDM}) as
\begin{equation}
  f_{\mathrm{DM}}(p,t) d\Omega= \zeta_{\rm RD/MD}  \times \frac{\xi \tilde f(x)}{ x^2}\,,
  \label{eq:fDMNinstfin}
\end{equation}
where the RD or MD dependent prefactor was defined in eq.~(\ref{eq:prefMDRD}). Notice that all the results of eqs.~(\ref{eq:RaMD}), (\ref{eq:RaRD}), and (\ref{eq:fDMNinstfin}) assume no changes in the number of relativistic dof available between $t_F$ and $t_\e$.

\section{The non-cold dark matter imprint}
\label{sec:lyman-alpha-n_eff}

The DM arising from PBH evaporation might be fast enough so as to
erase small scale structures. Constraints from such effects have
already been estimated in several previous works using different
methods, see e.g.~\cite{Fujita:2014hha, Lennon:2017tqq}. In this work,
for the first time, we extract NCDM constraints by making use of the
universal distributions obtained in Sec.~\ref{sec:extended-reheating} for
the non-instantaneous reheating case and calculate the resulting matter power spectrum.

The main constraints will arise from the Lyman-$\alpha$ forest flux
power spectrum, which probes hydrogen clouds at redshifts $2\lesssim z
\lesssim 6$, and has been shown to provide constraints on the matter
power spectrum on small scales~\cite{Ikeuchi1986, Rees1986}. However,
these scales are in the highly non-linear regime, and as such require
computationally expensive hydrodynamical N-body simulations, which are
currently only available for a limited subclass of
models~\cite{bolton17}. To circumvent the need for new N-body
simulations, in this paper we use {\sc class} to extract the
linear matter power spectrum of our NCDM scenario and the
corresponding transfer function, similar to what is usually done for
thermal warm dark matter (WDM) models~\cite{Viel:2005qj}. We will show that, even though
the PBH evaporation distribution function can differ from a thermal
WDM distribution (which would follow a FD
distribution, depicted by the blue dashed curve in Fig.~\ref{fig:spectr}), 
the resulting transfer functions have a form similar to those of thermal
WDM. In the same spirit as~\cite{Viel:2005qj} -- or more
recently~\cite{Murgia:2017lwo} -- we will fit the resulting transfer
functions with a minimal set of parameters, incorporated in the
breaking scale. This constitutes one of our main results, and
will allow us to extract Lyman-$\alpha$ constraints for our scenarios.

In addition, our NCDM can affect the effective number of relativistic
degrees of freedom at the time of CMB emission or BBN, see
also~\cite{Hooper:2019gtx,Hooper:2020evu,Masina:2020xhk}. Again,
making use of the distribution function of Sec.~\ref{sec:NCDM-PSD}, we
extract the contribution to the effective number of relativistic
non-photonic species, $\Delta N_{\mathrm{eff}}$, coming from the NCDM
from PBH in full generality, i.e. without assuming that the NCDM is
still relativistic at CMB or BBN time, following the approach
of~\cite{Merle:2015oja}. This treatment of $\Delta N_{\mathrm{eff}}$
differs from the default implementation for NCDM in {\sc class}, and
also from e.g.~\cite{Hooper:2019gtx,Masina:2020xhk}, the results of
which are valid assuming that the NCDM is relativistic at the CMB or
BBN time. Nevertheless, we reach the conclusion that the resulting
$\Delta N_{\mathrm{eff}}$ constraints cannot currently provide any
bound on the parameter space for a 2 dof fermionic DM.

\subsection{Estimate for the Lyman-$\alpha$ constraint}
\label{sec:estim-lyman-alpha}

Before going through the results arising from {\sc class}, we first
provide an estimate of the velocity constraints following an approach
similar to \cite{Fujita:2014hha,Masina:2020xhk}\footnote{Also see~\cite{Lennon:2017tqq} for a somewhat different approach, which takes greater account of the phase-space distribution, although without calculating the power spectrum, as we shall do later.}. For this purpose, we make use 
of the typical mean momenta at the time of evaporation obtained in
Sec.~\ref{sec:NCDM-PSD}. The mean velocity today of NCDM particles
arising from PBH evaporation (that have not yet virialised) should
satisfy
	\begin{equation}
	\langle v\rangle|_{t=t_0} = a_\e\times \frac{\langle p\rangle|_{t=\tau}}{m_{\mathrm{DM}}} 
			= \q\frac{\text{keV}}{m_{\mathrm{DM}}}\w \q\frac{M_{F}}{\mpl}\w^{1/2} \times
		\begin{dcases*}
		6.4 \times 10^{-7}  & for  $\beta<\beta_c$\,, \\
  		5.5 \times 10^{-7}   &  for   $\beta>\beta_c$\,,
		\end{dcases*}	
	\label{eq:vDMt0}
	\end{equation}
assuming that the DM momentum redshifts as $\propto 1/a$ from
the end of evaporation onwards.
In eq.~(\ref{eq:vDMt0}) we have used the scale
factor of eqs.~(\ref{eq:aRD}) and (\ref{eq:aMD}) assuming $T_F>T_{\rm EW}$, and the mean momentum
obtained in eq.~(\ref{eq:meanpEvapRDMD}) assuming non-instantaneous
RH. Notice that the mean DM velocity
today increases for increasing PBH mass, while the mean momentum at the
time of evaporation is smaller for higher PBH mass, see
eqs.~(\ref{eq:meanpEvap}) and~(\ref{eq:meanpEvapRDMD}). It is
important to take into account the redshifting of the momenta (which
scales with $a_\e\propto M_F^{3/2}$) at later times in order to get the
relevant DM velocity dependence in the PBH mass.

This velocity can be compared to the one expected for thermal WDM
particles that saturate the Lyman-$\alpha$ bound. The typical
velocity today of thermal DM particles that decoupled while still
being relativistic is estimated to be~\cite{Bode:2000gq}
\begin{equation}
  v_\W|_{t=t_0} \approx 3.9\times 10^{-8}\,\q\frac{\keV}{m_\W}\w^{4/3} \,.
  \label{eq:vWDM}
\end{equation}
Imposing that the mean velocity of eq.~(\ref{eq:vDMt0}) does not exceed the WDM velocity of eq.~(\ref{eq:vWDM}), for a WDM mass saturating the Lyman-$\alpha$ bound, $m^{{\rm Ly}-\alpha}_\W$, we get the constraint
	\begin{equation}
	m_{\mathrm{DM}} \gtrsim \q\frac{m^{{\rm Ly}-\alpha}_\W}{\keV}\w^{4/3}\q\frac{M_{F}}{\mpl}\w^{1/2} \times
	\begin{dcases*}
	 16 \, \keV  & for  $\beta<\beta_c$\,, \\
  	14 \, \keV   &  for   $\beta>\beta_c$\,.
	\end{dcases*}
	\label{eq:MDMconsFuj}
	\end{equation}
We will see in Sec.~\ref{sec:lyman-alpha-constr} that a dedicated analysis agrees with this simple estimate up to a factor of $\sim 3$. The dependence of
the limits on the NCDM mass from PBH evaporation in $m^{{\rm Ly}-\alpha}_\W$ and $M_F$ is also well recovered using {\sc class} together with the distribution functions of Sec.~\ref{sec:NCDM-PSD}.

\subsection{Lyman-$\alpha$ constraints from the transfer function}
\label{sec:lyman-alpha-constr}

We now make use of the non-instantaneous phase-space distributions
obtained in Sec.~\ref{sec:NCDM-PSD}, using them as an input for NCDM
in the Boltzmann code {\sc class}\footnote{Notice that in {\sc class},
  NCDM is assumed to be composed of a 2 dof species by default. As a
  result, we have implemented ${\tt f_0}= 2 \times f_{\rm DM}$, with
  ${\tt f_0}$ the NCDM phase-space distribution of {\sc class} and
  $f_{\rm DM}$ from eq.~(\ref{eq:fDMNinstfin}). In our simulations, we
  fix the parameter ${\tt deg_{ncdm} }$ of {\sc class}, counting the number
  of NCDM generations, to one. This amounts to one fermionic 2
  dof DM species, i.e. in our notations, to take $g_{\rm DM}=2$.} and
we extract the resulting matter power spectrum, assuming that the NCDM
accounts for all the DM. In order to parametrise the small scale
suppression of the matter power spectrum within a given NCDM model
with respect to the equivalent CDM case, one can express the ratio
between the CDM power spectrum, $P_{\rm{CDM}}(k)$, and the power
spectrum of some new DM species $X$, $P_{X}(k)$, in terms of
the transfer function $T_X$, defined as
\begin{equation}
P_{X}(k) = P_{\rm{CDM}}(k) \, T^2_{X}(k) \,,
\label{eq:pwdm}
\end{equation}
where $k$ is the wavenumber. It has been shown that the NCDM $T^2(k)$
can usually be parametrised in terms of a finite set of parameters and
physical inputs. In particular, in the thermal WDM case,
Refs.~\cite{Bode:2000gq,Viel:2005qj} use the following parametrisation to
describe the transfer function:
\begin{equation}
T_{X}(k) = \left(1+ (\alpha_{X} k)^{2\mu}\right)^{-5/\mu} \,,
\label{eq:twdm}
\end{equation}
where $\mu$ is a dimensionless exponent and $\alpha_X$ is the breaking
scale. A more general parametrisation that can be applied to a larger
set of NCDM models was also introduced in~\cite{Murgia:2017lwo,Murgia:2018now,Archidiacono:2019wdp}.

In the case of thermal WDM, Ref.~\cite{Viel:2005qj} obtained a very
good fit to the N-body simulations for $\mu=1.12$ and
\begin{eqnarray}
  	\alpha_{\rm{WDM}} 
		&=&  0.049\q \frac{m_{\rm WDM}}{1\,\text{keV}}\w^{-1.11}\q\frac{\Omega_{\rm WDM}}{0.25}\w^{0.11}\q\frac{h}{0.7}\w^{1.22}h^{-1}\text{Mpc}\,,\label{eq: alphaWDM}	\label{eq:alphWDM}
\end{eqnarray}
in terms of the WDM mass $m_{\rm WDM}$.  The analysis of
Ref.~\cite{Viel:2013fqw}, and more recently
\cite{Palanque-Delabrouille:2019iyz}, obtained a bound of
$m_\W>3.3\,$keV and $5.3\,$keV at $95\,\%$ C.L., respectively, from
Lyman-$\alpha$ flux observations. We note, however, that there are
claims in the literature~\cite{Baur:2017stq, Garzilli:2019qki} that
these $m_\W$ bounds depend heavily on the assumptions made about the
instantaneous temperature and pressure effects of the intergalactic
medium. Indeed, when relaxing these assumptions
\cite{Garzilli:2019qki} find a bound of $m_\W > 1.9\,$keV at $95\,\%$
C.L.. With this in mind, we will take the bound of
\begin{equation}
m_\W>3\,{\rm keV}  
\label{eq:Ly-WDM}
\end{equation}
as a conservative limit from Lyman-$\alpha$ on WDM, and the breaking
scale saturating this bound is $\alpha_{\W}=1.3\times 10^{-2}\,\text{Mpc}\,h^{-1}$. Further improvements on this bound might arise with extremely
large telescopes, see~\cite{Simon:2019kmm}.

We find that the transfer function of eq.~(\ref{eq:twdm}) with $\mu=1.12$
provides a very good fit to NCDM from PBH evaporation for a large set of
DM and PBH masses. The breaking scale $\alpha_{\rm PBH}$ in these cases will then be parametrised as
\begin{eqnarray}
	\alpha_{\rm PBH} = \q \frac{m_{\rm DM}}{1\,\text{eV}}\w^{-0.83}\q\frac{M_{\rm F}}{M_p}\w^{0.42} \times   \begin{dcases*}
  60.4 \, \text{Mpc}\,h^{-1} & if  $\beta<\beta_c$\,,\\
  53.2\, \text{Mpc}\,h^{-1}  &  if   $\beta>\beta_c$\,,
  \end{dcases*}
  \label{eq: alphaRD}
\end{eqnarray}
in the case of non-instantaneous reheating with $\Omega_{\rm DM}h^2=0.12$ today. We checked that the fit is valid for $
1.5\times 10^{-3}<\alpha_{\rm PBH}\times h/{\rm Mpc}<0.5$ with a
deviation of maximum 15\,\%. The same range of scales for $\alpha_\W$
would correspond to $m_\W \in [0.12,22]$\,keV, i.e. encapsulating
the Lyman-$\alpha$ bound of eq.~(\ref{eq:Ly-WDM}).  If we had
considered instantaneous RH, we would have a slightly larger breaking
scale with prefactors of 68.2 (59.7) for $\beta <\beta_c$ ($\beta
>\beta_c$). The dependency in $m_{\mathrm{DM}}$
and $M_F$ is inspired by the analytic result of
eq.~(\ref{eq:MDMconsFuj}). Indeed, imposing that the breaking scale in
the PBH case should always be smaller than the breaking scale of WDM
saturating the Lyman-$\alpha$ limit, we obtain a constraint on the
mass of DM arising from PBH evaporation given by
\begin{eqnarray}
  \label{eq:mDMCLASS}
m_{\mathrm{DM}} \geq \q \frac{m^{{\rm Ly}-\alpha}_\W}{\rm keV}\w^{4/3}\q\frac{M_F}{M_p}\w^{1/2} \times
  \begin{dcases*}
  5.2 \,  \mathrm{keV}  & if  $\beta<\beta_c$\,,\\
  4.4 \,  \mathrm{keV}  &  if   $\beta>\beta_c$\,.
  \end{dcases*}
  \end{eqnarray}
The general result of
eq.~(\ref{eq: alphaRD}) can be used to model the effect of DM from PBHs
on the matter power spectrum, and thus we can use this, together with eq.~(\ref{eq:mDMCLASS}), to extract constraints from any small
scale probe already testing the properties of thermal WDM.

\begin{figure}
        \centering
        \includegraphics[width=.8\linewidth]{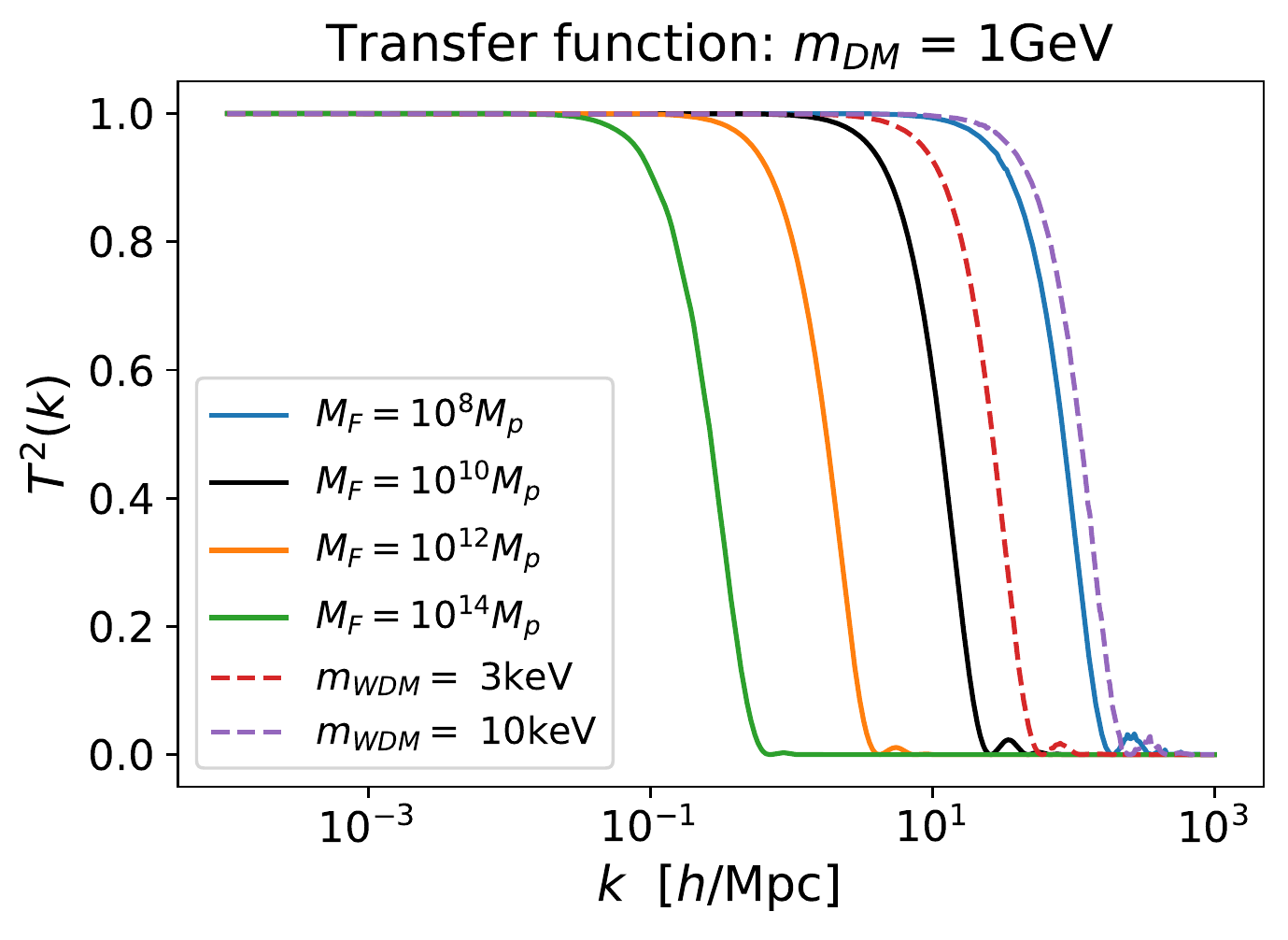}
        \caption{ Transfer functions for NCDM from PBH
          non-instantaneous evaporation with ${m_{\mathrm{DM}}= 1}$\,GeV and
          $\beta>\beta_c$ (continuous lines) and WDM with $m_\W= 3$
          and 10\,keV (red and purple dashed line respectively). The
          different coloured continuous lines correspond to different
          choices of PBH mass between $10^{14} M_p$ (left-most
          continuous green line) and $10^{8} M_p$ (right-most continuous
          blue line). In all cases, it is assumed that the DM accounts
          for $\Omega_\mathrm{DM} h^2=0.12$.}
        \label{fig:transferfn}
\end{figure}

In Fig.~\ref{fig:transferfn} we show some illustrative examples of
transfer functions. The transfer function for a WDM candidate with
$m_\W=3$\,keV, saturating our choice of conservative Lyman-$\alpha$ bound of
eq.~(\ref{eq:Ly-WDM}), and with $m_\W=10$\,keV, chosen as an illustrative example, are shown with dashed red and purple lines for comparison.  
In contrast, the continuous lines illustrate the
case of NCDM with $m_{\mathrm{DM}}= 1$\,GeV arising from PBH evaporation for PBH
masses between $10^{14} M_p$ (left-most continuous green line) and
$10^{8} M_p$ (right-most continuous blue line). In all cases, it has
been assumed that the NCDM accounts for $\Omega_\mathrm{DM} h^2=0.12$. The
continuous lines have been obtained by using the DM phase-space
distributions of Sec.~\ref{sec:ncdm-phase-space} assuming
non-instantaneous RH with $\beta>\beta_c$, see
Sec.~\ref{sec:extended-reheating}. We see that for fixed NCDM mass,
higher PBH mass imply higher velocities at the time of structure
formation, see eq.~(\ref{eq:vDMt0}) with $\langle v\rangle\propto
M_F^{1/2}$, and thus an exponential cut in the transfer function
arising at lower wave number $k$, or equivalently, at higher length
scale. It is already well visible from Fig.~\ref{fig:transferfn} that
a transfer function with similar parametrisation as the WDM will be
able to describe NCDM from PBH evaporation. As a result, we can
directly take away from Fig.~\ref{fig:transferfn} that, for $m_{\mathrm{DM}}=1$\,GeV, PBHs with mass $M_F<10^{10}M_p$ are not allowed by the
conservative Lyman-$\alpha$ bound of eq.~(\ref{eq:Ly-WDM}), while the
most stringent constraints from~\cite{Palanque-Delabrouille:2019iyz}
would extend the exclusion to masses slightly below $M_F\sim
10^{8}M_p$.

Let us emphasise that in Fig.~\ref{fig:transferfn}, we always consider
cases where the NCDM accounts for $100\,\%$ of the DM. The reason for
this is that in mixed NCDM+CDM models, the transfer function presents
a non-zero plateau at large $k$, which still contributes substantial power on small scales. As such, for these mixed cases the Lyman-$\alpha$ bounds need
to be adjusted and this requires a dedicated analysis~\cite{Boyarsky:2008xj,Baur:2017stq}. 
Indeed, in~\cite{Baur:2017stq} a generic formula is presented to deal with mixed WDM+CDM models; however, the shape of the resulting transfer functions for these cases is different to those we obtain in the case of the PBH sourced DM considered here. The reason for this is that the underlying mechanism by which the relic abundance is changed in these models is different, making a direct comparison non-trivial. A detailed analysis of how to reuse the bound presented in~\cite{Baur:2017stq} is beyond the scope of this work.
Thus, here we only apply the bounds from Lyman-$\alpha$ when the NCDM accounts
for all of the observed DM relic density today.

At this point, it is important to emphasise that in this paper 
we have considered only a monochromatic distribution of PBH masses. 
Nonetheless, in many circumstances the PBH mass function can be expected to spread over a large range of masses, see e.g. the
discussion in the recent review~\cite{Carr:2020xqk}. Even when assuming that the initial overdensity is monochromatic, the critical collapse phenomenon~\cite{Choptuik:1992jv,Niemeyer:1997mt, Musco:2008hv} gives rise to an extended mass function, which spreads towards lower
masses~\cite{Kuhnel:2015vtw}. Within the context of PBHs from inflation~\cite{Green:1999xm,Green:2016xgy} a log-normal distribution appears to fit a large class of scenarios (also see the discussion in~\cite{Kuhnel:2015vtw,Kuhnel:2017pwq,Carr:2018poi}). However, 
a general analysis for any non-monochromatic mass distribution is not possible. We therefore leave the derivation of limits in this scenario with non-monochromatic mass functions for future work.

\subsection{Contribution to $\Delta N_{\mathrm{eff}}$ }
\label{sec:delta-n_eff}

The DM arising from PBH evaporation can also significantly contribute to the
effective number of relativistic non-photonic species, $\Delta
N_{\mathrm{eff}}$, at the time of last scattering or BBN, provided the
DM particles are relativistic enough at those times. Current CMB data
tell us that $\Delta N_{\mathrm{eff}}(t_{\mathrm{CMB}})<0.28$ at 95\,\% C.L.,
using the latest measurements from the Planck collaboration (TT, TE,
EE+lowE+lensing+BAO). Notice that the present tension in the Hubble
constant measurement, can increase this upper bound to $\Delta
N_{\mathrm{eff}}(t_{\rm CMB})<0.52$ at 95\,\% C.L. (TT, TE,
EE+lowE+lensing+BAO+R18)~\cite{Aghanim:2018eyx}. Bounds of similar
order are expected to arise from BBN measurements, see
e.g.~\cite{Pitrou:2018cgg}, keeping in mind that those bounds can be
analysis-dependent, see e.g.~\cite{Aghanim:2018eyx,Schoneberg:2019wmt}
for a discussion. These
bounds are expected to improve by one order of magnitude with the
upcoming CMB Stage IV mission, which is predicted to have a
sensitivity of $\sigma \left( \Delta N_{\mathrm{eff}}(t_{\mathrm{CMB}}) \right)
\sim 0.06$~\cite{Abazajian:2019eic}.

In order to evaluate the contribution to $\Delta N_{\mathrm{eff}}$, we have to
carefully account for the fact that our DM is not always relativistic
at the time of interest. We follow the approach
of~\cite{Merle:2015oja}, see also e.g.~\cite{Baumholzer:2019twf}. 
The contribution from DM to $\Delta N_{\mathrm{eff}}(T)$ at a
given time, at SM radiation temperature $T$, is given by
\begin{eqnarray}
  \Delta N_{\mathrm{eff}} (T)&=&\frac{\rho_{\mathrm{DM}} (T)- m_{\mathrm{DM}}n_{\mathrm{DM}}(T)}{\rho_{rel \, \nu} (T)/N_{\mathrm{eff}}^{\nu}(T)}\cr
    &=&\frac{g_{\mathrm{DM}}}{2}\frac{30}{\pi^2}\frac87 \frac{\TN (T)^4}{T^4} \left(\frac{T}{T_\nu}\right)^{4} \, \zeta_{\rm RD/MD}\frac{m_{\mathrm{DM}}}{\TN(T) }   
\cr
    &&\times \int dx \left(\left(1+x^2\frac{\TN(T)^2 }{m_{\mathrm{DM}}^2}\right)^{1/2}-1\right) \xi \tilde f(x)\,,
\label{eq:Neff1}
\end{eqnarray}
 see App.~\ref{sec:entropy} for the details.
In the second equality the RD-MD dependent prefactor is given by
eq.~(\ref{eq:prefMDRD}). Notice that the ratio
$\left({T}/{T_\nu}\right)^{4}$ evolves between BBN and today from 1 to
$ \left({11}/{4}\right)^{4/3}$.  One can easily check that the
prefactor of $x^2$ in the integrand can be non-negligible for large
PBH mass and low DM mass, e.g. ${\TN(t_{\rm CMB}) }/{m_{\mathrm{DM}}}=1.32$ for
$m_{\mathrm{DM}}=10^{-3}$\,GeV and $M_F=10^{14}\times M_p$. In order to provide
simple estimates of $\Delta N_{\mathrm{eff}}(T)$, let us consider separately
the highly relativistic and non-relativistic cases, i.e. the cases
where the ratio $\TN(T)/m_{\mathrm{DM}}$ is much greater or smaller than one.
 
When $\TN(T)/m_{\mathrm{DM}}\gg 1$, the DM particles are still relativistic at the temperature
$T$ and eq.~(\ref{eq:Neff1}) simply reduces to
\begin{eqnarray}
  \Delta N_{\mathrm{eff}}^{\rm rel} (T)
  &=&\frac{120 \, \zeta_{\rm RD/MD}}{7 \pi^2} \left(\frac{ T_F \, a_\e}{T_{\nu} \, a(T)}\right)^4  \, \frac{N_{\mathrm{DM}}\langle p\rangle|_{t=\tau}}{T_F}  \label{eq:Neffrel}\\
  &&\cr
  &\simeq& \frac{g_{\mathrm{DM}}}{2} \begin{dcases*}
   1.2\times 10^{-1}\beta\times \frac{M_F}{M_p} & if  $\beta<\beta_c$\,,\\
     4.1\times 10^{-2} & if   $\beta>\beta_c$\,.
  \end{dcases*}
  \label{eq:NeffrelRDMD}
\end{eqnarray}
In the second equality we make use of the fact that the product
$({T}/{T_\nu})/(T\times a(T))$ is approximatively constant between
electron decoupling and today. We also use the results for $a_\e$,
$N_{\mathrm{DM}}$, and $\langle p\rangle|_{t=\tau}/T_F$ obtained in
the previous sections.  These results agree with the {\sc class}
outputs that assumes by default that the NCDM component is
relativistic. Furthermore, we find that in the PBH dominated case
before evaporation, i.e. $\beta > \beta_c$, the $ \Delta
N_{\mathrm{eff}}^{\rm rel}$ contribution from DM arising from PBH
evaporation does not depend on the PBH mass. This constant dependence
agrees with the results
of~\cite{Hooper:2019gtx,Masina:2020xhk}\footnote{We get similar
  results as in~\cite{Hooper:2019gtx} for $\beta>\beta_c$.  In the
  case $\beta<\beta_c$, we disagree in the PBH mass dependence
  of~\cite{Hooper:2019gtx} but we agree
  with~\cite{Masina:2020xhk}. However, we trust that the dependence in
  eq.~(\ref{eq:NeffrelRDMD}) is correct, as for $\beta=\beta_c$ we
  recover a mass independent result of the MD case. }.  Comparing the
$\beta> \beta_c$ results to the current bounds on $\Delta
N_{\mathrm{eff}}$, it appears that current data are at the limit of
restricting further the viable parameter space of a two dof fermionic
DM species evaporating from PBH. In addition, when $\beta< \beta_c$,
we typically have $\beta\times M_F/M_p< 0.3$ using
eq.~(\ref{eq:betac}). As a result, for any $\beta$, we do not expect
to get extra constraints on the viable DM parameter space making use
of $\Delta N_{\mathrm{eff}}$ constraints when considering
$g_{\mathrm{DM}}=2$. Current CMB experiments would be able to
  test $\Delta N_{\mathrm{eff}}$ for $\beta>\beta_c$ increasing the
  number of DM dof by a factor of $\sim 7$, while future CMB
  experiments would just need a minor change in dof to test $\Delta
  N_{\mathrm{eff}}$.

Clearly, if relativistic DM does not enhance sufficiently
$N_{\mathrm{eff}}$ to be constrained, we do not expect further bounds
in the non-relativistic case.  Indeed, when $\TN(T)/m_{\mathrm{DM}}\ll
1$, the $\Delta N_{\mathrm{eff}}$ is no longer given by
eq.~(\ref{eq:Neffrel}) and the correctly computed contribution will
always be smaller. Therefore, considering the relativistic result of
eq.~(\ref{eq:Neffrel}), as is the case in {\sc class} for example,
might result in an over-constraining value for $\Delta
N_{\mathrm{eff}}$.  Nonetheless, for the sake of the discussion, let
us briefly describe how the $\Delta N_{\mathrm{eff}}$ contribution
changes at a given time when the NCDM particles become non-relativistic.  When $\TN(T)/m_{\mathrm{DM}}\ll
1$, eq.~(\ref{eq:Neff1}) tends to
  \begin{eqnarray}
  \Delta N_{\mathrm{eff}}^{\rm NR} (T)
  &=&\frac{120 \, \zeta_{\rm RD/MD} }{7 \pi^2} \left(\frac{ T_F \,  a_\e}{T_{\nu} \, a(T)}\right)^4 \,  \frac{N_{\mathrm{DM}}\langle p^2\rangle|_{t=\tau}}{T_F^2} \, \left(\frac{ T_F a_\e}{2m_{\mathrm{DM}} a(T)}\right)\,.\label{eq:NeffNR}
  \end{eqnarray}
Due to the non-relativistic nature of DM, its kinetic energy is
expected to scale with the momentum squared $ p^2/2m_{\mathrm{DM}}\propto
1/a^2$ instead of $ p\propto 1/a$. As a result, we are left with
one negative power of the scale factor in the second line of
eq.~(\ref{eq:NeffNR}) that is not compensated by one negative power of
the temperature $T$. In the non-relativistic case, we thus expect $\Delta
N_{\mathrm{eff}} (T)$ to differ between BBN and CMB epoch. In particular, at
CMB time, we find:
\begin{eqnarray}
  N_{\mathrm{eff}}^{\rm NR}(T_{\rm CMB})  &=& \frac{g_{\mathrm{DM}}}{2} \frac{\langle p^2\rangle|_{t=\tau}}{T_F^2} \, \left( \frac{\rm GeV}{m_{\mathrm{DM}}}\right)\begin{dcases*}
   1.4\times 10^{-12}\beta \, \left( \frac{M_F}{M_p}\right)^{3/2} & if  $\beta<\beta_c$\,,\\
     8.6\times 10^{-13} \, \left( \frac{M_F}{M_p}\right)^{1/2} & if   $\beta>\beta_c$\,,
  \end{dcases*}
  \label{eq:NeffNRCMB}
\end{eqnarray}
where we expect ${\langle p^2\rangle|_{t=\tau}}/{T_F^2}$, defined
similarly as ${\langle p\rangle|_{t=\tau}}/{T_F}$, to be $\sim{\cal O}
(10)$. At BBN we would obtain a result a factor of $4\times 10^6$ 
larger (for $T_{BBN}=4$\,MeV) if the particles were still
non-relativistic at this earlier time. Even for the largest PBH masses
and lowest DM masses considered here, the resulting $\Delta
N_{\mathrm{eff}}^{\rm NR}$ is always left unconstrained.

\section{Viable parameter space from cosmology}
\label{sec:viable-param-space}

\begin{figure}[t]
	\centering
	{\includegraphics[scale=0.5]{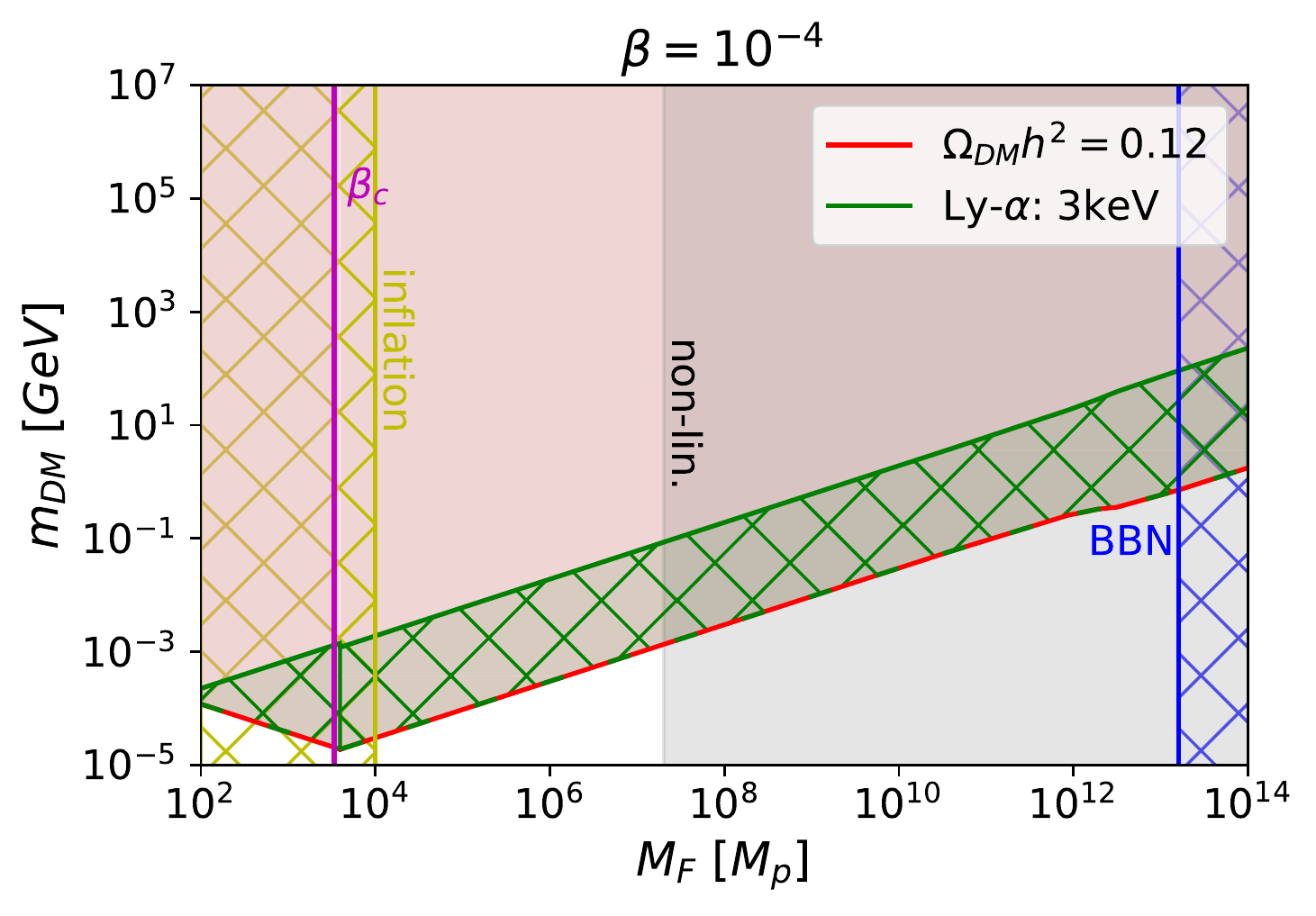}}
	\quad
	    {\includegraphics[scale=0.5]{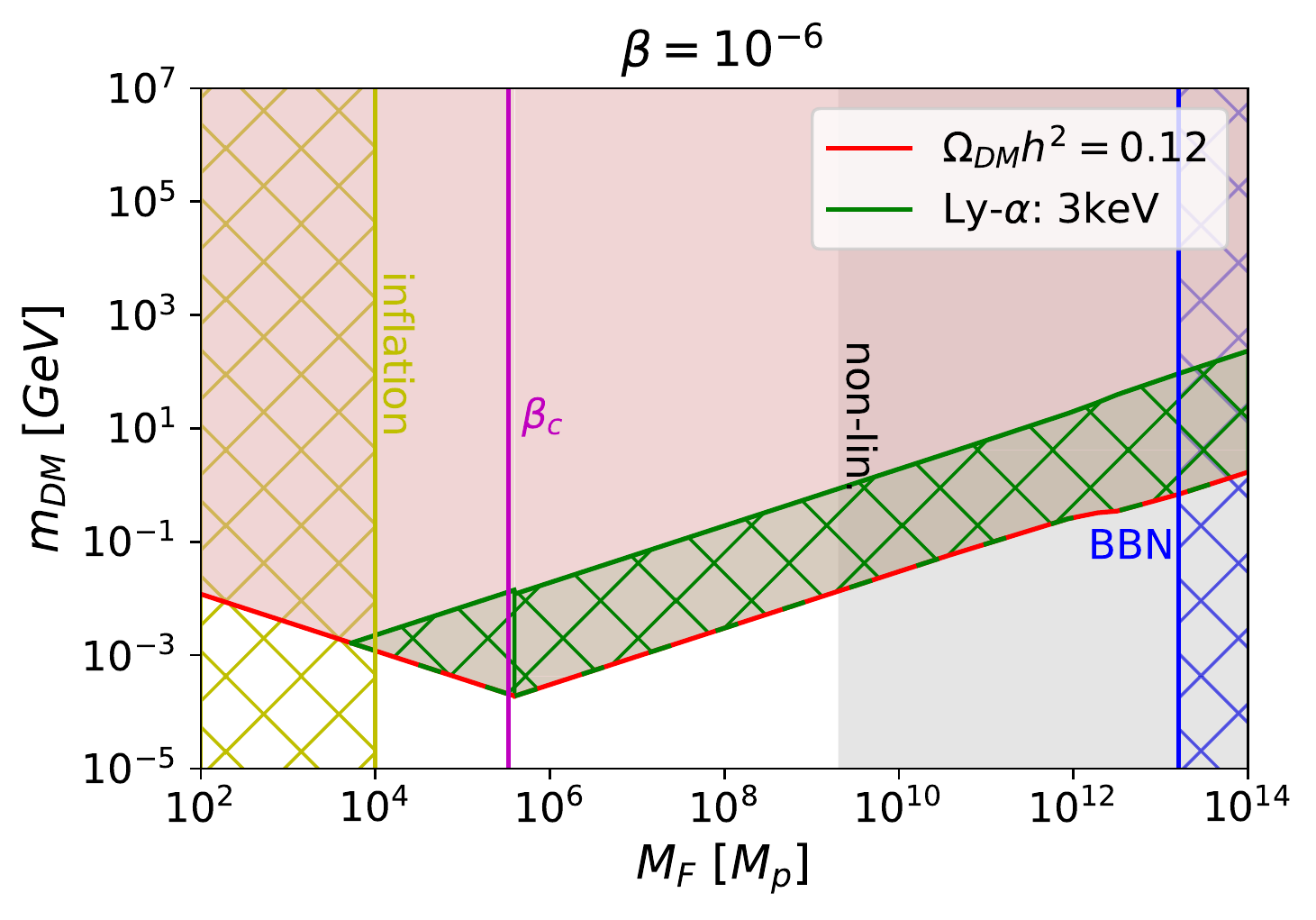}}\\
           	{\includegraphics[scale=0.5]{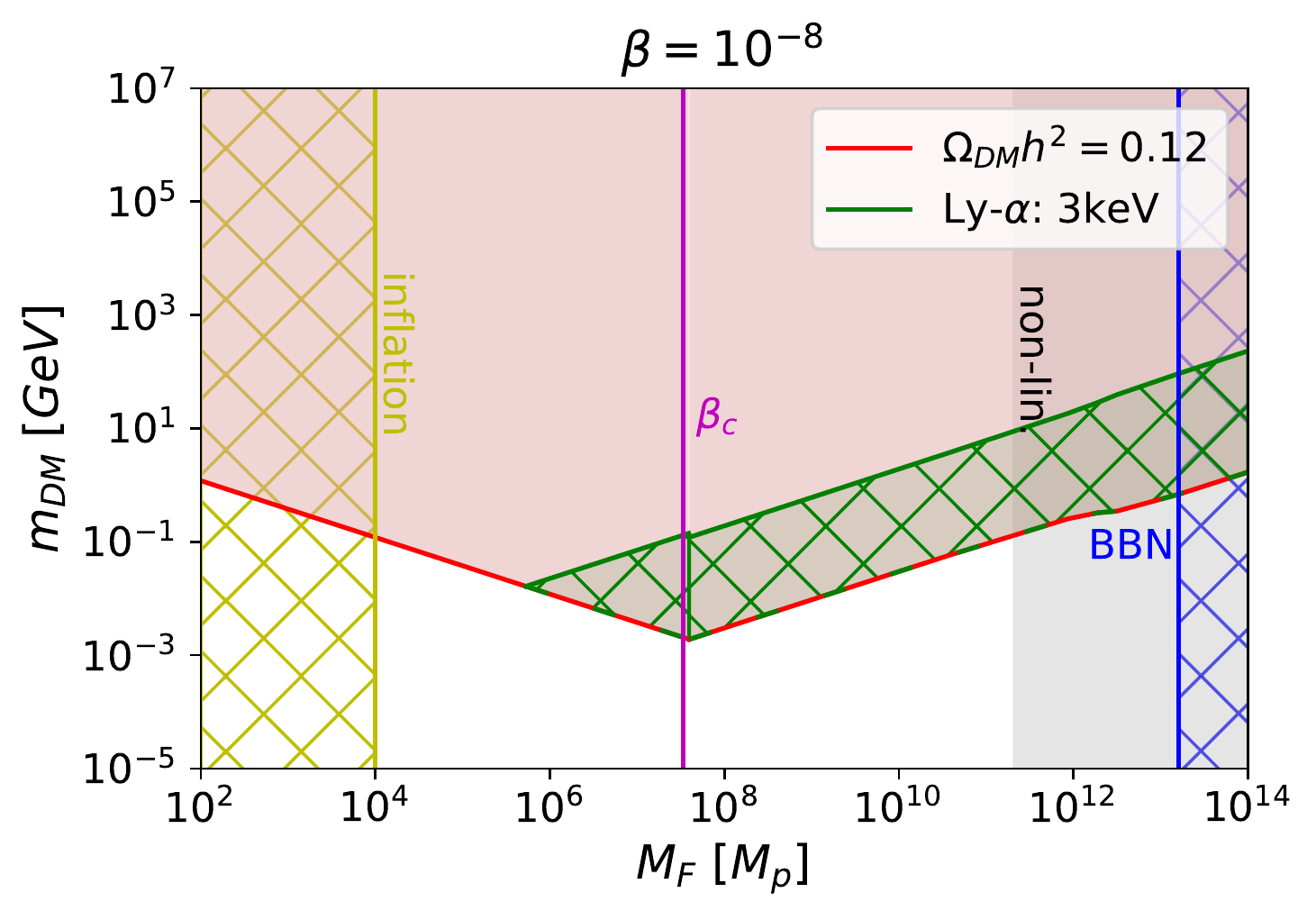}}
	\quad
	{\includegraphics[scale=0.5]{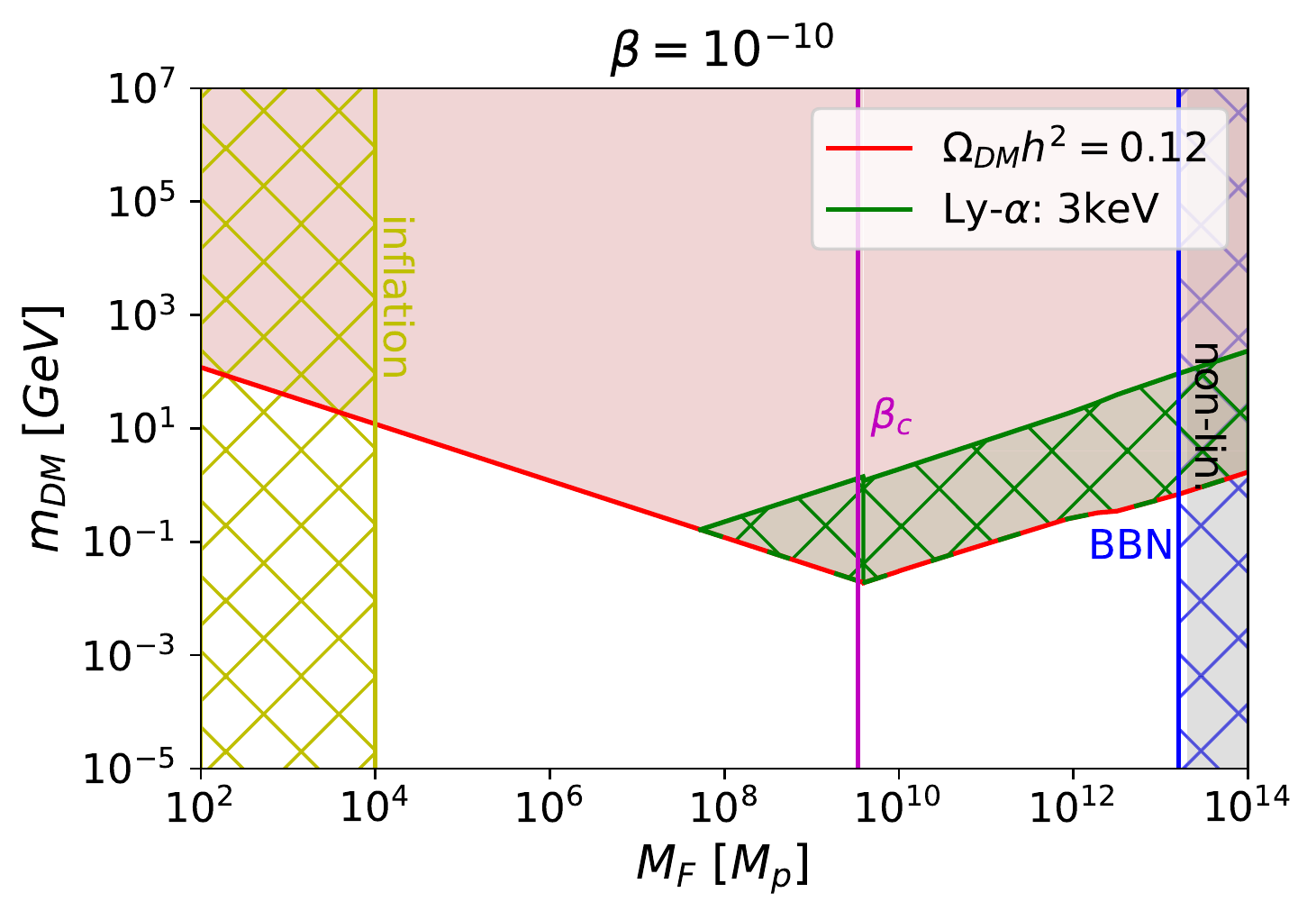}}
	\caption{ Viable parameter space for NCDM arising from PBH
          evaporation, for different values of $\beta$.  The models on
          the red line produce the correct DM abundance today, while
          in the red region DM overcloses the universe. The blue,
          yellow, and green hatched zones are excluded by BBN,
          inflation, and Lyman-$\alpha$ constraints respectively, including where they overlap with the red line. In
          the grey region we do not fully trust our results as extra compact
          objects might have formed in the early MD era. The white
          regions show the viable parameter space that avoids all
          constraints.}
	\label{fig:viab}
\end{figure}

We can now summarise all the results obtained in the previous
sections. For this purpose we show in Fig.~\ref{fig:viab} the
viable parameter space in the plane $(M_F,m_{\mathrm{DM}})$ for NCDM arising
from PBH evaporation, for four different choices of the initial PBH
relative abundance $\beta$. The initial black hole mass viable
parameter space is bounded from above by the BBN constraint (blue hatched
region), see eq.~(\ref{eq:BBN}), and from below by inflation (yellow
hatched region), see eq.~(\ref{infl}). In the grey region we do
not fully trust our results for DM production, as overdensities could
have become non-linear and additional long-lived massive objects might
have formed, see eq.~(\ref{eq:limnon-lin}).

The red line shows us where NCDM obtained from evaporation saturates
the CMB bound on the DM relic abundance $\Omega_{\mathrm{DM}}
h^2=0.12$, in agreement with the {\sc class} result. Above the red line
(light red region), PBHs produce too much DM and overclose
the universe. The $\Omega_{\mathrm{DM}} h^2=0.12$ line changes slope
for $\beta=\beta_c$, which is indicated with a magenta vertical
line. On the left (right) of this line, the universe is RD (MD) before
evaporation, i.e. $\beta<\beta_c$ ($\beta>\beta_c$) and the DM mass
should decrease (increase) with increasing PBH mass in order to
account for all the DM, as expected from the analytic results of
eq.~(\ref{eq:OmDM2}). 

Notice that for PBH masses $M_F\gtrsim
  10^{10} M_p$ the temperature of the bath at evaporation drops below
  $T_ {\rm EW}$. As a result, the number of SM relativistic dof at evaporation decreases: $g_{*}^{SM}(t_\e)<106.75$, and the
  scale factor at evaporation will be affected, see
  eq.~(\ref{eq:RaRD}). The detailed impact of a change in the number
  of relativistic dof on the NCDM relic density is provided in
  App.~\ref{ap: changing the DOF}. The largest change happens for
  $\beta>\beta_c$, in which case $\Omega_{\rm DM}\propto
  (g_{*}(t_\e))^{3/4}/g_{*s}(t_\e)$, see
  eq.~(\ref{eq:Om-srcaling}). For the largest PBH masses considered
  here, we expect $g_{*}^{SM}(t_\e)$ to get as small as 10.75, implying
  a relic density larger by a factor of $\sim 2$ at most.  As
  a result, for fixed $M_F$, $m_{\rm DM}$ will be reduced by up to a
  factor 2, explaining why the red curves scale down from $M_F\gtrsim
  10^{10} M_p$. 

The green line indicates where the Lyman-$\alpha$ bound, corresponding
to a WDM of 3\,keV, would exclude NCDM from PBH evaporation if this
NCDM accounts for $\Omega_\mathrm{DM} h^2=0.12$. We see that the Lyman-$\alpha$
bound only crosses the relic abundance line for $\beta<\beta_c$. Now
any couple of $(m_{\mathrm{DM}},M_F)$ giving rise to all the DM --
i.e. any point on the red line -- lying below the Lyman-$\alpha$ line
would be excluded by any other bound on WDM excluding $m_\W<3$\,keV. As
a result, the Lyman-$\alpha$ bound excludes all scenarios giving rise
to all the DM between the green line and the red line (green hatched
region). Notice that NCDM with masses as large as 100\,GeV can be
excluded by this Lyman-$\alpha$ bound when arising from PBH
evaporation. 

For $M_F\gtrsim 10^{10} M_{p}$ the green line
  scales slightly  up. This is again due to the change in the number of
  relativistic dof at evaporation that affect the scale factor $a_\e$,
  scaling as $g_{*}(t_\e)^{1/4}/ g_{*s}(t_\e)^{1/3}$, entering in the
  determination of our Lyman-$\alpha$ bound. Going to lower PBH
  masses, we get lower $g_{*}(t_\e)$ and slightly higher
  $a_\e$. Therefore, we expect from the analytic estimate of
  Sec.~\ref{sec:estim-lyman-alpha} that the NCDM velocity today will
  increase and the Lyman-$\alpha$ bound will be slightly reinforced. This is
  indeed observed in our simulations, see App.~\ref{ap: changing the
    DOF}. In particular, we use  eq.~(\ref{eq:mDMCLASSFull}) in our plots of Fig.~\ref{fig:viab}. 

Below the red line the Lyman-$\alpha$ bounds are no
longer valid, as they were obtained assuming $\Omega_{\rm DM}
h^2=0.12$. If instead we consider that only part of the DM is made of
NCDM, the form of the transfer function, and thus the
Lyman-$\alpha$ bounds, drastically change. As discussed before, 
getting the Lyman-$\alpha$ constraints in this region is non-trivial, see \cite{Baur:2017stq}, and is beyond the scope of this paper. 
As such, we leave the region below the red line unconstrained, giving the
possibility to account for part of the DM as NCDM from PBH evaporation. 

In conclusion, we see that NCDM from PBH evaporation can still account for a subcomponent of the DM for any $\beta$. By combining eqs.~(\ref{eq:OmDM2}) and (\ref{eq:mDMCLASS}), we find the NCDM from PBH evaporation can account for all
the DM when
	\begin{equation}
	  \beta < 0.016 \, \beta_c.
          \label{eq:omDM+lym}
	\end{equation}
        For the lightest possible $M_{F}$ this corresponds to $\beta\lesssim 5 \times 10^{-7}$ and $m_{\mathrm{DM}}\gtrsim$ 2\,MeV. For larger $M_{F}$ the allowed $\beta$ ($m_{\mathrm{DM}}$) decreases (increases). Such viable scenarios appear on the red line above the green region in the two bottom plots of Fig.~\ref{fig:viab}.

\section{Leptogenesis from PBH evaporation}
\label{sec:lepto-from-pbh}

Some words regarding baryogenesis are now in order. Although it has been speculated since the 1970s that PBHs may themselves be the seed for baryogenesis~\cite{Carr:1976zz,Zeldovich:1976vw,Toussaint:1978br,Dolgov:1979mz,Turner:1979zj,Turner:1979bt,Dolgov:1980gk,Barrow:1990he,Majumdar:1995yr,Upadhyay:1999vk,Dolgov:2000ht,Bugaev:2001xr,Baumann:2007yr,Hook:2014mla,Banks:2015xsa,Fujita:2014hha,Hamada:2016jnq,Morrison:2018xla,Carr:2019hud,Garcia-Bellido:2019vlf}, we will limit ourselves to demanding the compatibility of our scenario with baryogenesis. If we assume our scenario produces the complete DM relic abundance, then we require eq.~(\ref{eq:omDM+lym}) to be satisfied. The PBHs  produce a negligible entropy dump, and hence the scenario is entirely compatible with baryogenesis in general.

For example, let us consider the standard leptogenesis scenario in greater detail, assuming eq.~(\ref{eq:omDM+lym}) is satisfied. The heavy Majorana neutrinos will happily coexist in the plasma together with the PBHs and produce the baryon asymmetry through their CP violating decays in the standard way~\cite{Buchmuller:2005eh,Davidson:2008bu}. Similarly, the PBH decay will only be affected in a negligible way by the emission of the Majorana neutrinos. Furthermore, it is possible to show that the neutrinos produced by the PBH will only contribute a negligible amount to the baryon asymmetry, outside of a tiny sliver of parameter space. 

Let us examine the last point in greater detail\footnote{Similar arguments have been made in~\cite{Fujita:2014hha}, for $\beta > \beta_{c}$, and more generally in~\cite{Morrison:2018xla}, which differs by not restricting the CP violation as we shall do.}. Let us assume a hierarchical spectrum of heavy neutrinos and impose the Davidson-Ibarra bound on the CP violation in the decays of the heavy eigenstates~\cite{Davidson:2002qv}. Namely, the usual CP-violating parameter is bounded by
	\begin{equation}
	\epsilon \lesssim \frac{ 3M_{N}\delta m_{\nu} }{ 8\pi v_{\phi}^{2} }\,,
	\end{equation}
where $v_{\phi} = 246$\,GeV is the electroweak VEV, $M_{N}$ is the heavy neutrino mass, and finally ${\delta m_{\nu} \sim 0.05}$\,eV is maximum difference in light neutrino masses. Although this constraint is usually applied to the lightest of the heavy mass eigenstates, the form of the CP violation~\cite{Covi:1996wh} means similar arguments also hold for the heavier states.

\begin{figure}[t]
\begin{center}
\includegraphics[scale=0.8]{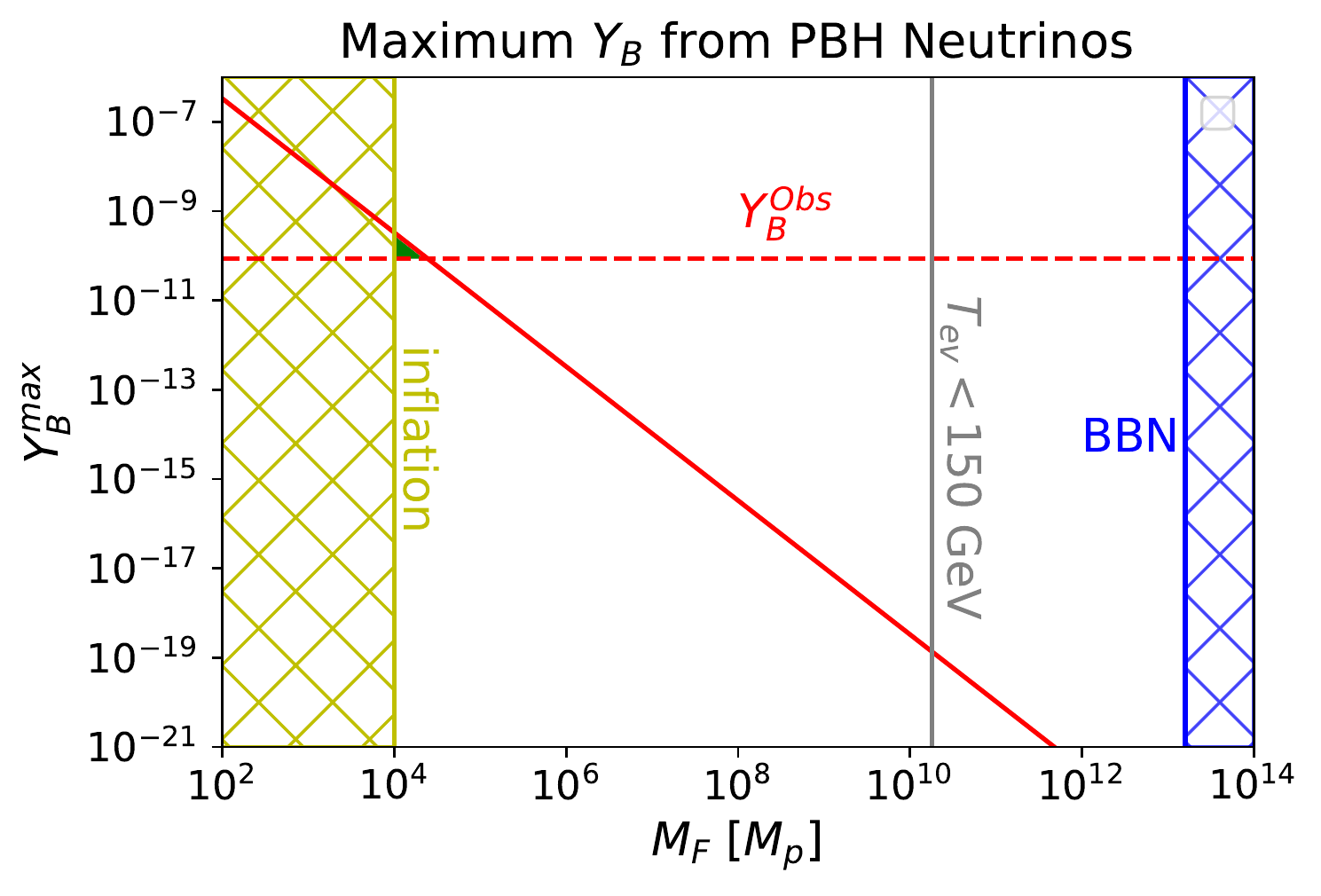}
\end{center}
\vspace*{-5mm}
\caption{The red continuous line shows the maximum yield of baryons from decays of heavy neutrinos
  produced by PBHs in the standard leptogenesis scenario. This assumes that the NCDM produced by the
  PBH accounts for $\Omega_\mathrm{DM} h^2=0.12$ and does not spoil Lyman-$\alpha$
  constraints, see
  eq.~(\ref{eq:omDM+lym}). The observed asymmetry is shown as a dashed
  line. A further suppression, not taken into account here, occurs to
  the right of the grey vertical line due to the sphalerons switching
  off before the PBHs fully decay. The region of parameter space where
  the CP asymmetry could be accounted for is highlighted in green.}
\label{fig:lgen}
\end{figure}

To quantify, we can write the baryon yield from the heavy neutrinos emitted by the PBHs as~\cite{Buchmuller:2005eh,Davidson:2008bu}
	\begin{equation}
	Y_{B} = \epsilon \kappa N_N Y_{\rm BH} \,,
	\end{equation}
where $\kappa \simeq 28/79$ is the sphaleron reprocessing factor, $N_N$ is the number of neutrinos emitted per PBH, and $Y_{\rm BH} \equiv n_{\rm BH}/s$ is the number density of PBHs normalised to the entropy density. The process is inherently out-of-equilibrium as $T_{\rm BH}$ exceeds the surrounding plasma temperature during the PBH decay process, thus satisfying the third Sakharov condition~\cite{Toussaint:1978br,Dolgov:1979mz}. The numerator of $Y_{\rm BH}$ at formation time, $t_F$, is given in eq.~\eqref{eq:nBHtF}. The denominator follows immediately from eq.~\eqref{eq:plasmaTatform}. In short $Y_{\rm BH} \simeq 0.06 \, \beta \, (M_{p}/M_{F})^{3/2}$. Finally we require $N_N$ which can be found using eq.~\eqref{eq:Nj} assuming $M_N < T_F$. Note as $\epsilon \propto M_{N}$ and if $M_{N} > T_{F}$ the number $N_N \propto (M_{p}/M_{N})^{2}$~\cite{Fujita:2014hha}, the maximum $Y_{B}$ occurs when $M_{N} = T_{F}$, which we now impose. Realistically, for $M_{N} = T_{F}$ there is already a suppression in $N_N$, but an approximation is made here in order to find a conservative bound on $Y_{B}$. Combining all of the above, together with our Lyman-$\alpha$ constraint of eq.~(\ref{eq:omDM+lym}) for the DM arising from the PBH account for $\Omega h^2=0.12$, yields a limit 
	\begin{equation}
	Y_{B} < 3.3 \times 10^{-4} \left( \frac{ \delta m_{\nu} }{ 0.05\, \mathrm{eV} } \right) \left( \frac{ M_p }{ M_F } \right)^{3/2}\,.
	\end{equation}
Comparing this with the observed asymmetry $Y^{\rm Obs}_{B} \simeq 0.86 \times 10^{-10}$~\cite{Aghanim:2018eyx}, in Fig.~\ref{fig:lgen} we see that the PBH sourced neutrinos can contribute sizeably to the asymmetry only in a tiny area of the parameter space highlighted in green, close to the bound coming from inflation. We re-emphasise that the heavy neutrinos in the bath can be more plentiful and hence still source the baryon asymmetry in the rest of the parameter space. Of course, in alternative scenarios the amount of CP violation can also be raised, which broadens the parameter space for which PBHs play a crucial role for baryogenesis.

\section{Conclusions}
\label{sec:concl}

In this paper we have revisited the case of NCDM particles arising from PBH evaporation. We do not assume any interaction for the DM
except for gravitational interactions, and we focus on the production of DM particles with mass below the BH temperature at formation, $m_{\mathrm{DM}}<T_F$. Such NCDM, not even feebly coupled to the SM, can leave a testable imprint on the cosmology by suppressing small scale structures, see e.g.~\cite{Fujita:2014hha,Allahverdi:2017sks,Lennon:2017tqq,Morrison:2018xla,Hooper:2019gtx,Hooper:2020evu,Masina:2020xhk} for previous analysis in this direction. We consider a Dirac-delta PBH mass distribution, having formed at the end of inflation in a radiation dominated era, with a proportion $\beta$ normalised to the critical density. Depending on whether $\beta$ is smaller or larger than some critical value $\beta_c$, the PBHs will evaporate in a radiation dominated or PBH dominated era. As already well-known, this directly affects the predictions for the NCDM relic abundance. Furthermore, PBH evaporation as such is already constrained by the allowed scale of inflation and should not spoil BBN. This constrains the BH mass at formation, $M_F$, to be in the range of ${[10^4,2\times 10^{13}]\times M_p}$. 

Concerning the NCDM, we have fully accounted for the fact that the production at evaporation does not happen instantaneously. We have extracted the DM phase-space distribution and interfaced it with the public Boltzmann solver {\sc class}. We recover the DM relic abundance and find agreement with analytic estimates, independently of the instantaneous or non-instantaneous nature of the evaporation. In addition, we have updated the constraints arising from Lyman-$\alpha$ flux measurements. In particular, we have obtained a fit to the transfer functions arising from multiple simulations with {\sc class}, parametrised in terms of a single free parameter: the breaking scale $\alpha_{PBH}$. The resulting fitting function appears to follow the very same parametrisation as the one for WDM in~\cite{Viel:2005qj},
even though the velocity distribution arising from PBH evaporation
differs from a thermal distribution, displaying a higher velocity tail
and a peak squeezed to lower velocities, see
Fig.~\ref{fig:spectr}. The breaking scale $\alpha_{PBH}$ depends on
$M_F$ and $m_{\mathrm{DM}}$, as specified in eq.~(\ref{eq: alphaRD}). Imposing
that $\alpha_{PBH}$ is smaller than the WDM breaking scale saturating
the Lyman-$ \alpha$ bounds gives us a generic constraint on the DM
mass as a function of $M_F$ in eq.~(\ref{eq:mDMCLASS}). The resulting
dependence agrees with analytic estimates we derived here, see also~\cite{Fujita:2014hha,Masina:2020xhk}.

We also computed the NCDM contribution to the number of non-photonic
relativistic degrees of freedom $\Delta N_{\mathrm{eff}}$ in full
generality, i.e. not assuming beforehand that the DM is relativistic
contrarily to e.g.~{\sc class}, and provided simple analytic estimates
for the extreme cases. We conclude that $\Delta N_{\mathrm{eff}}$ does
not further constrain the viable parameter space for our choice of
NCDM with two fermionic dof. Future CMB missions will
be on the verge of  testing such contribution to $\Delta N_{\mathrm{eff}}$
while increasing the number of DM dof by a factor of
seven, current CMB constraints would already test $\Delta
N_{\mathrm{eff}}$.
 
Our findings are summarised in the plots of Fig.~\ref{fig:viab} in the planes of $(M_F,m_{\mathrm{DM}})$ for four different values of $\beta$. More generally, if a monochromatic distribution of PBHs came to dominate the energy density of the universe we find they cannot have decayed into light DM, $m_{\mathrm{DM}}<T_F$, accounting for the complete relic abundance. On the other hand, if there is some other source of DM, we cannot quantify with the current method the fraction that could be NCDM from PBH evaporation, given that we can only apply the Lyman-$\alpha$ constraints to $\Omega_{\rm DM}h^2=0.12$. We can also account for all the DM imposing a conservative Lyman-$\alpha$ bound corresponding to a $m_\W=3$\,keV when satisfying eq.~(\ref{eq:omDM+lym}). Imposing the inflation bound, this implies that $\beta\lesssim 5 \times 10^{-7}$ and $m_{\mathrm{DM}}\gtrsim$ 2\,MeV. Notice that NCDM with mass as large as 100\,GeV can be excluded by this Lyman-$\alpha$ bound when arising from PBH evaporation.

For the sake of completeness, we also revisit the possibility to
  account for both all the DM and for leptogenesis from heavy
  neutrinos arising form PBH. Taking into account the Lyman-$\alpha$
  constraints derived here, maximising the baryon yield considering
  $M_N= T_F$, the PBH sourced neutrinos can only contribute sizeably
  to the asymmetry in a tiny area, close to the bound coming from
  inflation.

\section*{Acknowledgements}
We would like to thank Sebastien Clesse for insightful discussion and Matteo Lucca for feedback on the draft. All
the authors are supported by the Fonds National de la Recherche
Scientifique (FRS-FNRS).  IB is postdoctoral researcher of the
FRS-FNRS with the project ``\emph{Exploring new facets of DM}.'' QD, DCH,
and LLH are (partially) supported by the FNRS research grant number
F.4520.19.  QD benefits from the
support of the French community of Belgium through funding of a FRIA
grant. LLH is a Research Associate of the FNRS and is also
supported by the Vrije Universiteit Brussel through the Strategic
Research Program \textit{High-Energy Physics}.

\appendix
\section{Greybody factors }
\label{sec:greybody-factors}

We refer to the absorption probabilities $\Gamma_j(E,M_{\mathrm{BH}})$, the
coefficients entering in the evaluation of the number of particles per
unit time and energy interval of eq.~(\ref{eq:dNdtdE}), as greybody
factors. As mentioned in the text, these coefficients tend to the
geometrical-optics limit -- used in the bulk of the text -- at high
energy, but fall off more rapidly at low energy and this fall is spin-dependent. These greybody factors affect the power
emitted by a black hole $-dM_{\mathrm{BH}}/dt\sim \int dE E \Gamma_j$ and the
number of emitted particles $dN/dt\sim \int dE \Gamma_j$ {\it differently}. They are
also expected to affect the velocity distribution, shifting the maximum
velocity peak to higher velocity, leading to a slight underestimation
of the total portion of relativistic particles, as already
underlined in~\cite{Lennon:2017tqq}.

When considering the geometrical-optics limit to compute the rate of
BH mass loss, we obtain the result of eq.~(\ref{eq:dMdt}) involving a
number of relativistic dof
\begin{equation}
  g_{* \rm{BH}}=\sum_b g_b+\sum_f \frac78 g_f \,,
  \label{eq:gstarBH}
\end{equation}
where $g_{j=b,f}$ counts all the bosonic and fermionic dof with mass
smaller than the temperature of the BH. As mentioned in
Sec.~\ref{sec:black-holes-evap}, accounting for all SM
relativistic dof plus an extra two-component fermionic DM
particle gives $g_{* \rm{BH}}=106.75+2\times 7/8= 108.5$ and $e_T=7.6\times
10^{-3}$.

The detailed computation of
this mass loss rate, including the full treatment of the greybody
factors, was obtained
in~\cite{PhysRevD.41.3052,PhysRevD.44.376}. Following this
prescription, the $e_T$ factor of eq.~(\ref{eq:eT}) would become
\begin{equation}
  \tilde e_T={\cal G} \frac{  \tilde g_{* \mathrm{BH}}}{30720\pi} \quad{\rm with} \quad \tilde g_{* \mathrm{BH}}=\sum_i g_j X_j\,,
  \label{eq:eTMcGibbon}
\end{equation}
for particles much lighter than the BH temperature, where ${\cal G}\simeq 3.8$, $g_j$
denotes the number of dof and $X_j$ encapsulates the spin-dependent
greybody factor effect, see
e.g.~\cite{Hooper:2019gtx,Masina:2020xhk}. In particular, a particle
of spin 0, spin 1/2, or spin 1 would have $X_j=1.82, 1.0, 0.41$
respectively. Considering again the emission of all the SM dof and 2
dof fermionic DM, Ref.~\cite{Hooper:2019gtx} gets $\tilde g_{*
  \mathrm{BH}}=110$ and $ \tilde e_T=4.3\times 10^{-3}$. The
similarity between the numerical values of $g_{* \mathrm{BH}}$ and
$\tilde g_{* \mathrm{BH}}$ is to be expected, as the SM is mainly made
of fermions for which $X_i=1.0$ in eq.~(\ref{eq:eTMcGibbon}), while
they are weighted by $7/8$ in eq.~(\ref{eq:gstarBH}). Overall,
neglecting the detailed effect of the greybody factors in computing
the BH mass loss rate, we overestimate  the latter by a factor
$\sim 2$, as $ e_T$ is approximatively twice as large as $\tilde
e_T$. This also implies that the BH lifetime of eq.~(\ref{eq:tau}) is
underestimated by a factor of 2.  On the other hand, the number of
emitted particles of species $j$ is expected to scale as
\begin{eqnarray}
  \tilde N_j&=&g_j X'_j \frac{ 81\zeta(3)}{4096 \pi^4 {\tilde e_T}}\frac{M_F^2}{M_p^2}\, ,
\label{eq:tildNj}
\end{eqnarray}
where the $X'_j$ coefficient has yet
to be determined and is expected to differ from $X_j$ of
eq.~(\ref{eq:eTMcGibbon}), as $N_j$ and $dM_{\mathrm{BH}}/dt$ arise from
different energy-dependent integrants involving $\Gamma_j$.

In the previous
works evaluating the DM production from PBH evaporation, the detailed
greybody factor impact on the emitted DM number of particles has been
rather diverse, sometimes being omitted or partially taken into
account, see
e.g. refs.~\cite{Fujita:2014hha,Allahverdi:2017sks,Lennon:2017tqq,Baumann:2007yr,Morrison:2018xla,Hooper:2019gtx,Masina:2020xhk,
  Hooper:2020evu}.  Here for simplicity -- and in order to provide a self-consistent
analysis -- we have chosen to use the geometrical-optics limit for the
mass loss, the DM number density, and the impact on small scale
structure. We can, however, easily estimate how a change from $e_T$ to
$\tilde e_T$ would affect the NCDM relic abundance modulo the
uncertainty on $X'_j$ in eq.~(\ref{eq:tildNj}). Indeed, the abundance
scales as: $\Omega_{\rm DM}\propto N_{\rm DM}\times \zeta_{\rm
  RD/MD}\times a_\e^3$, see eq.~(\ref{eq:OmDM1}), while the number of
DM particles from a BH, the scale factor at evaporation, and the
prefactor $\zeta_{\rm RD/MD}$ show the following dependencies in
$e_T$: $ N_{\rm DM}\propto e_T^{-1}$, $a_\e\propto e_T^{-1/2}$ and
$\zeta_{\rm RD/MD}\propto e_T^{3/2}$ for $\beta<\beta_c$ or
$\zeta_{\rm RD/MD}\propto e_T^{2}$ for $\beta>\beta_c$, see
eqs.~(\ref{eq:Nj}), (\ref{eq:aRD}) and~(\ref{eq:aMD}),
and~(\ref{eq:prefMDRD}). As a result, a full treatment of the greybody
factors will give rise to a NCDM relic density
\begin{eqnarray}
\label{eq:Omtilde}
\tilde\Omega_{\mathrm{DM}} (t_0)=\Omega_{\mathrm{DM}} (t_0) \times X'_{\rm DM} \times
\begin{dcases*}
e_T/\tilde e_T  & if  $\beta<\beta_c$\,,\\
(e_T/\tilde e_T)^{1/2} &  if   $\beta>\beta_c$\,.
\end{dcases*}
\end{eqnarray}
where $X'_{\rm DM}$ is the prefactor $X'_j$ in eq.~(\ref{eq:tildNj}) for
$j=$ DM that would need to be computed, and $\Omega_{\mathrm{DM}}
(t_0)$ is the relic abundance in the geometrical-optics limit. For
$T_\e>T_{\rm EW}$, we thus expect $\tilde\Omega_{\mathrm{DM}} (t_0)$ to differ from $\Omega_{\mathrm{DM}} (t_0)$ by a factor $1.8\times X'_{\rm DM}$ for $\beta<\beta_c$ and a factor $1.3\times X'_{\rm DM}$ for $\beta>\beta_c$. 

The impact of the full treatment of the greybody factors on the
Lyman-$\alpha$ constraints depends both on the change in $e_T$, which
affects $a_\e$, and on the expected shift of the peak in momentum of $d
N(p)/dp$ to higher momenta, which would affect $\langle
p\rangle|_{t=\tau}$. Here we can only derive the impact of the change
in $e_T$ on the breaking scale. Replacing $e_T$ with $\tilde e_T$, we
get the fitting formula
\begin{eqnarray}
  \alpha_{\rm PBH} = \q \frac{m_{\rm DM}}{1\,\text{eV}}\w^{-0.83}\q\frac{M_{\rm F}}{M_p}\w^{0.42} \times   \begin{dcases*}
    74.7 \, \text{Mpc}\,h^{-1} & if  $\beta<\beta_c$\,,\\
    65.8 \, \text{Mpc}\,h^{-1}  &  if   $\beta>\beta_c$\,.
\end{dcases*}
\label{eq: alphaRDHooper}
\end{eqnarray}
This would imply a strengthening of the bounds obtained in
Sec.~\ref{sec:lyman-alpha-constr} by $\sim 25\,\%$. The  shift
in the peak velocity to higher velocities would strengthen this bound
even further. 

\section{Validity of instantaneous evaporation approximation}
\label{sec:valid-inst-evap}
In this appendix we derive under which conditions the instantaneous
evaporation approximation is valid. Instantaneous evaporation takes
place when
\begin{equation}
	B^2 = \frac{H_F^{-2}}{\tau^2} \gg 1\,,
\end{equation}
with $H(t_F)$ the Hubble rate at formation \cite{Lennon:2017tqq}. It can easily be shown that in the BH production setup studied in this paper, the Hubble time at production is given by
\begin{equation}
	H_F^{-1} = \frac{2 M_F}{\gamma \mpl^2}\,.
\end{equation}
A such, we obtain
\begin{equation}
	B^2 = \frac{36 e^2_T}{\gamma^2} \q\frac{\mpl^4}{M_F^4}\w \leq \frac{36 e^2_T}{\gamma^2} \frac{1}{10^{16}}\,,
\end{equation}
where we used the bound coming from inflation in eq.~\eqref{infl}. Therefore, for the approximation to be valid we need to have $g_{\rm * BH} \gtrsim 10^{10}$.

\section{Changing the number of degrees of freedom} 
\label{ap: changing the DOF}
In this appendix we generalise our results to take into account changing numbers of dof during the evaporation process. Note that here, as well as in the bulk of the text, we work in the limit were the radiation density is dominated by SM dof after PBH evaporation, i.e. $g_{\rm DM} \ll g_{* \rm BH}$. If $g_{\rm DM}$ becomes sizable all the expressions below will be modified.

If the universe stays radiation dominated during evaporation, the ratio between the scale factor at formation and evaporation is given by
\begin{equation}
\frac{a_{F}}{a_\e} = \q\frac{g_{* s}(t_\e)}{g_{* s}(t_{F})}\w^{1/3}\q\frac{g_{*}(t_{F})}{g_{* }(t_\e)}\w^{1/4} \q\frac{3\epsilon_T}{\gamma}\w^{1/2} \q\frac{\mpl}{M_F}\w \qquad {\rm if }\quad \beta<\beta_c \,,
\label{eq:RaRDFull}
\end{equation}
following the same reasoning leading to eq.~\eqref{eq:RaRD}. It is clear that the above reduces to eq.~\eqref{eq:RaRD} when the number of dof remains fixed.

In contrast, it turns out that the analogous ratio for $\beta <\beta_c$, eq.~\eqref{eq:RaMD}, is not affected by a change in dof, at least at the level of our approximation. To see this, we first require the entropy boost factor, $D_{s}$, due to PBH decay. At formation, the PBH density normalised to entropy is given by
	\begin{equation}
	Y_{\rm BH}(t_{F}) \equiv \frac{ n_{\rm BH}(t_{F}) }{ s(t_{F}) } =  \beta \frac{  \rho_{R}(t_{F}) }{ M_{F} s(t_{F}) }\,.
	\end{equation}	
Precisely at the instant of evaporation, when the PBHs have reheated the thermal bath, the same factor is given by
	\begin{equation}
	Y_{\rm BH}(t_\e) = \frac{ \rho_{R}(t_\e) }{  M_{F} s(t_\e) }\,,
	\end{equation}
which implies
	\begin{equation}
	D_{s} = \frac{ Y_{\rm BH}(t_{F}) }{ Y_{\rm BH}(t_\e) } = \beta \frac{ g_{*}(t_{F}) }{ g_{*}(t_\e) } \frac{ g_{*s}(t_\e) }{ g_{*s}(t_{F}) }
	 \frac{ T(t_{F}) }{ T_\e } \,.
	\end{equation}
The ratio of scale factors can then be written as
	\begin{align}
	\frac{ a_{F} }{ a_\e } & = \left[  \frac{1}{D_{s}} \frac{ s(t_\e) }{ s(t_{F}) } \right]^{1/3}
	 = \left[ \frac{1}{\beta}  \frac{\rho_{R}(t_\e) }{\rho_{R}(t_{F})}\right]^{1/3}  = \left(\frac{ 16 e_T^2}{\gamma^2\beta}\frac{M_p^4}{M_F^4}\right)^{1/3}\,.
	\end{align}
To understand the result, note $\rho_{R}(t_{F})$ can be written in terms of $M_{F}$, through eq.~\eqref{eq:rhof}, in which the dof do not enter. Similarly, $\rho_{R}(t_\e)$ is set by the Hubble scale at decay, and hence the PBH lifetime $\tau$. Therefore, it depends on $g_{\rm * BH}$ but not on the relativistic dof of the bath.

Taking these results into account, the prefactors of eq.~\eqref{eq:prefMDRD}, appearing in various quantities, become
\begin{equation}
\zeta_{\rm RD} \to  \q\frac{g_{* s}(t_\e)}{g_{* s}(t_{F})}\w\q\frac{g_{*}(t_{F})}{g_{* }(t_\e)}\w^{3/4} \zeta_{\rm RD}\,, \qquad \qquad \zeta_{\rm MD} \to \zeta_{\rm MD}\,. 
\label{eq:prefMDRDFull}
\end{equation}
To understand the effect of changing the dof on the relic density we use eqs.~\eqref{eq:Nj}, \eqref{eq:omdm0}, \eqref{eq:aMD}, \eqref{eq:RaRDFull}, and \eqref{eq:prefMDRDFull}. This gives the following scaling
\begin{equation}
\Omega_{\rm DM}(t_0) \propto \begin{dcases*} 
\frac{ g_{\rm DM} \times g_{*}(t_{F})^{3/4} }{ g_{* \rm BH} \times g_{* s}(t_{F}) } & if  $\beta<\beta_c$\,,\\
\frac{g_{\rm DM} \times g_{*}(t_\e) ^{3/4}}{ g^{1/2}_{* \rm BH} \times g_{* s}(t_\e)}  & if  $\beta>\beta_c$\,.
\end{dcases*}
\label{eq:Om-srcaling}
\end{equation}
A first interesting thing we notice from this is that in the case of $\beta <\beta_c$ the relic density is not influenced by the dof at evaporation, while in the case of $\beta >\beta_c$ the dof at formation do not play a role. The results here are used in making the plots in Sec.~\ref{sec:viable-param-space}, which result in small features in the contours at points where the SM dof change substantially.

As the scale factor at evaporation is altered by changing the dof, so
is the free-streaming suppression.  Keeping $g_{*s} = g_{*}$, the
effects of a change in $g_{*}$ and in $g_{*{\rm BH} }$ are captured by
the following extra terms in the fitting formula for the breaking
scale:
\begin{eqnarray}
\alpha_{\rm PBH} = \q\frac{g_{\rm * BH}}{108.5}\w^{-0.42} \q\frac{g_{*}(t_\e)}{108.5}\w^{-0.07} \q \frac{m_{\rm DM}}{1\,\text{eV}}\w^{-0.83}\q\frac{M_{\rm F}}{M_p}\w^{0.42}
	\nonumber \\
 \times   \begin{dcases*}
60.4 \, \text{Mpc}\,h^{-1} & if  $\beta<\beta_c$\,,\\
53.2 \, \text{Mpc}\,h^{-1}  &  if   $\beta>\beta_c$\,.
\end{dcases*}
\label{eq: alphaRDFull}
\end{eqnarray}
This implies that \eqref{eq:mDMCLASS} becomes
\begin{eqnarray}
\label{eq:mDMCLASSFull}
m_{\mathrm{DM}} \geq \q\frac{g_{\rm * BH}}{108.5}\w^{-1/2} \q\frac{g_{*}(t_\e)}{108.5}\w^{-1/12} \q \frac{m^{{\rm Ly}-\alpha}_\W}{\rm keV}\w^{4/3}\q\frac{M_F}{M_p}\w^{1/2}
	\nonumber \\
 \times
\begin{dcases*}
5.2 \,  \mathrm{keV}  & if  $\beta<\beta_c$\,,\\
4.4 \,  \mathrm{keV}  &  if   $\beta>\beta_c$\,.
\end{dcases*}
\end{eqnarray}
The above dependence is inspired by our analytic estimates of Sec.~\ref{sec:lyman-alpha-n_eff} as the DM velocity is directly proportional to  $a_\e\propto g_{\rm * BH}^{-1/2} \q g_{*}(t_\e)\w^{-1/12}$. Also notice that the effect of $g_{*}(t_\e)$ is so suppressed by a small power that a large change in $g_{*}(t_\e)$ would be needed to see a significant effect (above our error margin in the fits).

\section{Radiation contributions}
\label{sec:entropy}
The entropy is given by
\begin{eqnarray}
  s&=& \frac{2\pi^2}{45}g_{*s} T^3 \quad {\rm with}\quad g_{*s}=\sum_b g_b \left(\frac{T_b}{T}\right)^{3} +\sum_f \frac78 g_f \left(\frac{T_f}{T}\right)^{3} \,,
\label{eq:s}
\end{eqnarray}
where $f$ and $b$ count all the fermionic and bosonic relativistic
dof and account for their temperature $T_f,\ T_b$
relative to that of the thermal bath, $T$. The energy density of the
relativistic particles is given by
\begin{eqnarray}
  \rho_R&=& \frac{\pi^2}{30}g_{*} T^4 \quad {\rm and}\quad g_{*}=\sum_b g_b \left(\frac{T_b}{T}\right)^{4} +\sum_f \frac78 g_f \left(\frac{T_f}{T}\right)^{4}\,.  
\label{eq:rhoR}
\end{eqnarray}
Also notice that at $T\lesssim$ MeV the neutrinos are decoupled while still
being non-relativistic. Using entropy conservation we can see that $ \frac{T_\nu}{T}= \left(\frac{4}{11}\right)^{1/3}$ for $T\ll$ MeV.
When the temperature of the bath of photons $T$ drops below $\sim$ MeV,
i.e. after $e^+e^-$ decoupling, we use 
\begin{equation}
  g_{*}(T)= 2 \left(1+\frac{7}{8}\left(\frac{T_\nu}{T}\right)^{4} N_{\mathrm{eff}}(T)\right)\quad [T\lesssim {\rm MeV}]\,.
  \label{eq:gstarNeff}
\end{equation}
In particular, around BBN we assume
${T_\nu}/{T}=1$, while at the time of last scattering we have
${T_\nu}/{T}=({4}/{11})^{1/3}$. The first term in
eq.~(\ref{eq:gstarNeff}) accounts for the photons, while the second term
accounts for the SM neutrinos and all other extra species that would
still be relativistic today. Focusing on the SM only, one gets for $T<m_e$
\begin{equation}
  g_{*,0}^{\rm SM}\simeq 2+6\frac{7}{8}\left(\frac{4}{11}\right)^{4/3}=3.36 \quad {\rm and }\quad  g_{* s,0}^{\rm SM}\simeq 2+6 \, \frac{7}{8}\left(\frac{4}{11}\right)=3.91 \,,
  \label{eq:gstar0}
\end{equation}
which implies the entropy today is $s_0=2.22\times 10^{-38}\,{\rm GeV^3}$ using $T_0= 2.7255$\,K. It is useful to remember that actually $N^{\nu}_{\rm eff}=3.046$ instead
of $N_\nu/2=3$ for $T\ll m_e$, due to spectral distortions in the
neutrino distribution function associated to non-instantaneous
decoupling and flavour oscillations~\cite{Mangano:2005cc}.

At this point, we can also define the contribution to $N_{\mathrm{eff}}$ from
new dark degrees of freedom at a given temperature $T$. From eq.~(\ref{eq:gstarNeff}), we have that the dark
degrees of freedom contribute as
\begin{eqnarray}
 \Delta N_{\mathrm{eff}} (T)&=&\frac{\sum_D g_{*D} \left(\frac{T_D}{T}\right)^{4}}{2\times \frac{7}{8}\left(\frac{T_\nu}{T}\right)^{4}} = \frac{\rho_{\rm rel}^D(T)}{\rho_{\rm rel}^{SM \nu}(T)/N_{\mathrm{eff}}^{\nu}}=\frac{\rho_D (T)-\sum_D m_Dn_D(T)}{\rho_{\rm rel}^{\rm SM \nu}(T)/N_{\mathrm{eff}}^{\nu}}\,,
  \label{eq:NeffD}
\end{eqnarray}
i.e. $N_{\mathrm{eff}}=N_{\mathrm{eff}}^{\nu}+\Delta
N_{\mathrm{eff}}$. In the first line $g_{\rm *D}= g_D$ for a boson and
$g_{\rm *D}= 7/8 \times g_D$ for a fermion, and $g_D$ counts particle and
antiparticle dof. In particular, for the case considered here,
$g_{\rm *DM}=\frac78 \times 2$ when the DM is relativistic. In
the third equality we have interpreted the relativistic contribution
of the dark species as their full contribution to the energy density
$\rho\propto\int d^3p E f_D$, with $E^2=p^2 +m_D^2$ minus the rest
frame contribution, with $n_D\propto \int d^3p f_D$, where $f_D$ is
the dark species phase-space distribution,
see~\cite{Merle:2015oja,Baumholzer:2019twf}. 

Finally, we note that the
contribution from neutrinos at a given time is given by
\begin{equation}
  \rho_{\rm rel}^{\rm SM \nu}(T)=2\times \frac78 \times N^{\nu}_{\rm eff} \frac{\pi^2}{30} T^4 \left(\frac{T_\nu}{T}\right)^{4}\,.
\end{equation}

\bibliographystyle{JHEP}
\bibliography{bibPBH}


\end{document}